\RequirePackage{algpseudocode}
\documentclass[final]{siamart190516}
\usepackage{amssymb}
\usepackage{graphicx, color}
\usepackage{enumitem}
\usepackage{tikz}
\usepackage{tkz-euclide}
\usetikzlibrary{positioning}
\usetikzlibrary{arrows.meta}
\usetikzlibrary {decorations.pathmorphing, decorations.text}

\usepackage[left=1in,right=1in,top=1in,bottom=1in,foot=0.5in]{geometry}
\usepackage{hyperref}
\usepackage{xcolor}
\hypersetup{
  colorlinks,
  linkcolor={red!70!black},
  citecolor={blue!50!black},
  urlcolor={blue!80!black}
}
\usepackage{algorithm}
\usepackage{algpseudocode}
\usepackage{cleveref}

\newcommand{\op}{\ensuremath{{\rm op}}}
\newcommand{\<}{\lessdot}
\newcommand{\osR}{\ensuremath{{{\rm cR}}}}
\newcommand{\sosR}{\ensuremath{{{\rm scR}}}}
\newcommand{\qosR}{\ensuremath{{{\rm qcR}}}}
\newcommand{\sqosR}{\ensuremath{{{\rm sqcR}}}}
\newcommand{\I}{J^{\circ}}
\newcommand{\J}{J}
\newcommand\mdoubleplus{\ensuremath{\mathbin{+\mkern-7mu+}}}
\renewcommand*{\append}{\mdoubleplus}

\newcommand*{\arc}[1]{X^\beta_{#1}}

\newcommand{\call}[1]{\ensuremath{\normalfont\textsc{#1}}}
\newcommand{\Let}{\State {\bfseries let} }

\renewcommand{\Return}{\State {\bfseries return} }
\algrenewcommand\algorithmicrequire{\textbf{Input:}}
\algrenewcommand\algorithmicensure{\textbf{Output:}}

\usepackage{bm}
\newcommand{\BD}{\bm{d}}

\crefname{claim}{claim}{claims}
\Crefname{claim}{Claim}{Claims}
\crefname{line}{line}{lines}
\Crefname{line}{Line}{Lines}
\crefname{enumi}{}{}
\Crefname{enumi}{}{}
\crefname{ineq}{inequality}{inequalities}
\Crefname{ineq}{Inequality}{Inequalities}
\creflabelformat{ineq}{#2{\upshape(#1)}#3}

\tikzset{vertex/.style={
  fill=black,
  draw=black,
  circle,
  inner sep=1pt,
  minimum size = 3pt
  }
}

\title{A simple and optimal algorithm for strict circular seriation }

\author{Mikhael Carmona\footnotemark[1] \footnotemark[2]
\and Victor Chepoi\thanks{LIS, Aix-Marseille Université, CNRS, and
Université de Toulon, Marseille, France (\email{\{mikhael.carmona, victor.chepoi, guyslain.naves,
pascal.prea\}@lis-lab.fr})}
\and Guyslain Naves\footnotemark[1]
\and Pascal Préa\footnotemark[1] \thanks{École Centrale de Marseille, Marseille, France}
}

\headers{A simple and optimal algorithm for strict circular seriation}{M. Carmona, V. Chepoi, G. Naves and P. Préa}

\date{}

\begin{document}

\maketitle

\begin{abstract}
 Recently, Armstrong, Guzm\'{a}n, and Sing Long (2021), presented an
 optimal $O(n^2)$ time algorithm for strict circular seriation
 (called also the recognition of strict quasi-circular Robinson
 spaces). In this paper, we give a very simple $O(n\log n)$ time
 algorithm for computing a compatible circular order for strict
 circular seriation. When the input space is not known to be strict
 quasi-circular Robinson, our algorithm is complemented by an
 $O(n^2)$ time verification of compatibility of the returned order.
 This algorithm also works for recognition of other types of strict
 circular Robinson spaces known in the literature. We also prove that
 the circular Robinson dissimilarities (which are defined by the
 existence of compatible orders on one of the two arcs between each
 pair of points) are exactly the pre-circular Robinson
 dissimilarities (which are defined by a four-point condition).
\end{abstract}



\section{Introduction} \label{s:intro}

A major issue in classification and data analysis is to visualize
simple geometrical and relational structures between objects based on
their pairwise distances. The classical {\it (linear) seriation problem}
(called also the matrix reordering problem), introduced by Robinson
\cite{Robinson}, asks to find a simultaneous ordering (or permutation)
of the rows and the columns of the dissimilarity matrix so that its
values increase monotonically in the rows and the columns when moving
away from the main diagonal in both directions. The permutation which
leads to a matrix with such a property is called a {\it compatible
order} and dissimilarity matrices admitting a compatible order are
called \emph{Robinson matrices}. The Robinson matrices can be thus
characterized by the existence of a compatible order $<$ and the
3-point condition $\BD(x,z)\ge \max\{ \BD(x,y), \BD(y,z)\}$ for any three
points $x,y,z$ such that $x<y<z$. If this inequality is strict, then
such a matrix is called \emph{strict Robinson}. A natural generalization
of Robinson dissimilarities and compatible
orders is to consider a circular order instead of a linear one. This
is often referred to as the \emph{circular seriation problem}.
Seriation (linear or circular) has numerous applications in data science, originating from various
research areas: archeological dating, hypertext orderings, overlapping clustering, gene expression,
DNA sequencing, DNA replication and 3D conformation, planar tomographic
reconstruction, quadratic assignment problem, numerical ecology, sparse matrix ordering,
musicology, matrix visualization methods, etc.


Due to its importance, 
the algorithmic problem of recognizing Robinson dissimilarities/matrices attracted the
interest of many authors. The existing recognition algorithms can be
classified into \emph{combinatorial} and \emph{spectral}. All
combinatorial algorithms use the correspondence between Robinson
dissimilarities and interval hypergraphs. The main difficulty arising
in recognition algorithms is the existence of several compatible
orders. The first recognition algorithm by Mirkin and Rodin
\cite{MiRo} consists in testing if the hypergraph of balls is an
interval hypergraph and runs in $O(n^4)$ time. Chepoi and Fichet
\cite{ChepoiFichet} gave a simple divide-and-conquer algorithm running
in $O(n^{3})$ time. Seston \cite{Seston} presented another $O(n^{3})$ time algorithm, by
using threshold graphs. In \cite{SestonThese}, he improved the
complexity of his algorithm to $O(n^2 \log n)$ by using a sorting of the
data and PQ-trees. Finally, in 2014, Préa and Fortin \cite{Prea} presented an
algorithm running in optimal $O(n^2)$ time. The efficiency of the
algorithm of \cite{Prea} is due to the use of the PQ-trees of Booth
and Lueker \cite{BoothLueker} as a data structure for encoding all
compatible orders. Even if optimal, the algorithm of \cite{Prea} is
far from being simple. Subsequently, two new recognition algorithms
were proposed by Laurent and Seminaroti: in \cite{LaSe1} they
presented an algorithm of complexity $O(\alpha\cdot n)$ based on
classical LexBFS traversal and divide-and-conquer (where $\alpha$ is
the depth of the recursion tree, which is at most the number of
distinct elements of the input matrix), and in \cite{LaSe2} they
presented an $O(n^2\log n)$ algorithm, which extends LexBFS to
weighted matrices and is used as a multisweep traversal. More
recently, in \cite{CaChNaPr} we gave a simple and practical $O(n^2)$
divide-and-conquer algorithm based on decompositions of dissimilarity
spaces into mmodules (subsets of points not distinguishable from the outside).
The spectral approach was introduced by Atkins
et al. \cite{AtBoHe} and was subsequently used in numerous papers. The
method is based on the computation of the second smallest eigenvalue
and its eigenvector of the Laplacian of a similarity matrix and
also uses PQ-trees to represent the compatible orders. The case when
the eigenvector is not simple was considered in the recent paper by
Concas et al. \cite{CoFeRoVa}.

The circular seriation problem takes its origins in the papers by Hubert,
Arabie, and Meulman \cite{Hub,HuArMe,HuArMe2}. 
More recently, circular seriation found interesting applications in planar tomographic
reconstruction, gene expression, DNA replication and 3D
conformation, see the papers \cite{CoShSiSi,LiLoXiWaCh,LiuLiYaNo}.
At the difference of the classical
seriation, where the notion of Robinson dissimilarity is a
well-established standard, in circular seriation there are several
non-equivalent notions of circular Robinson dissimilarities.  Hubert et al. 
\cite{Hub,HuArMe,HuArMe2} defined a class of circular
Robinson dissimilarities via a certain 4-point condition.  Brucker and Osswald
\cite{BrOs} undertaken a systematic study of various definitions of
circular Robinson dissimilarities from the point of view of
classification and combinatorics. In Robinson dissimilarity spaces,
the sets of balls, of 2-balls (intersections of two balls), and of clusters (maximal
cliques in the threshold graphs) are all interval hypergraphs.
Hypercycles, introduced and investigated by Quillot \cite{Qu}, are the
circular analogs of interval hypergraphs and are the hypergraphs whose
hyperedges can be represented as circular intervals (arcs). In the case of
circular Robinson dissimilarities, requiring that the ball, the
2-ball, or the cluster hypergraphs are hypercycles lead to three
different classes of dissimilarity spaces. Their structural properties have
been thoroughly studied by Brucker and Osswald \cite{BrOs}. The dissimilarities whose ball
hypergraph is a hypercycle is the most general one and was
characterized in \cite{BrOs} via a simple 4-point condition. 
We call such dissimilarities quasi-circular Robinson. To characterize
the dissimilarities for which the 2-ball or the cluster hypergraphs are hypercycles, 
Brucker and Osswald \cite{BrOs} introduced the notion of pre-circular
Robinson dissimilarities.

The algorithmic problem of recognizing circular Robinson dissimilarities was less studied
and optimal or even subcubic time algorithms for different versions of this problem
are not known. A spectral approach to circular seriation was developed in the papers
\cite{EvBrMuGo,IsGiVe,ReKeD'A}. Hsu and McConnell \cite{HsuMcCo} showed how to efficiently
recognize the hypercycles, by using a data structure, which they call PC-tree, and
which is a generalization of well-known PQ-trees. Using this result, Brucker and Osswald \cite{BrOs}
designed cubic time algorithms for recognizing quasi-circular 
Robinson dissimilarities. It comes to some surprise when
recently Armstrong, Guzm\'{a}n, and Sing Long \cite{ArGuSiLo} presented an optimal $O(n^2)$ time algorithm for
the recognition of strict quasi-circular Robinson dissimilarities.
Among other tools, their algorithm uses PQ-trees.

In this paper, we give a very simple $O(n \log n)$ time algorithm
which builds a compatible circular order for %
all versions of strict circular Robinson
dissimilarities, introduced and investigated in the papers
\cite{ArGuSiLo,BrOs,HuArMe}.
Then the adjunction of a verification step gives an optimal $O(n^2)$ time %
algorithm to recognize these dissimilarities. %
Our second main result is proving that
the pre-circular Robinson dissimilarities are exactly the
dissimilarities for which there exists a circular order $\<$ such that
for each pair $(x,y)$, the restriction of $d$ to one of the two arcs
between $x$ and $y$ is a Robinson dissimilarity (in the usual sense)
and $\<$ is its compatible order. To our knowledge, prior to our work
no results of this kind for circular seriation were known. Our result
shows that in fact pre-circular Robinson spaces should be called
circular Robinson spaces. 
Finally, the simplicity of our algorithm led us to other structural properties of
strict circular Robinson spaces, in particular we show that they admit only one or two
compatible circular orders and their opposites. The results of \cite{ArGuSiLo} and 
of this paper can be viewed as the first step toward designing efficient algorithms for circular seriation.
Designing algorithms which solve the circular seriation problem
in subcubic time (say, in $O(n^2\log n)$ or in optimal time $O(n^2)$) is an interesting \emph{open question}.
While a random circular Robinson space is almost surely a strict circular Robinson space,  circular Robinson
dissimilarities are an ideal case and the dissimilarity matrices, which are measured only
approximatively,  fail to satisfy the circular Robinson property. Therefore, the \emph{second open algorithmic problem} is the approximation of a
dissimilarity space by a circular Robinson space. In the linear case, 
these fitting problems
are NP-hard for 
$\ell_1$-norm~\cite{BaBr} and $\ell_\infty$-norm~\cite{ChFiSe} (no polynomial time algorithm is known for the $\ell_p$-norms with $1<p<\infty$)
and a constant factor approximation algorithm for $\ell_{\infty}$-fitting problem was designed in \cite{ChSe}.

The remaining part of the paper is organized as follows. In Section \ref{s:prel} we
define the various notions of circular Robinson spaces. In Section \ref{s:circ}
we prove our first main result about the characterization of pre-circular Robinson spaces. In Section \ref{s:prop} we
present the main properties of quasi-circular and strict circular Robinson spaces. We also show how to efficiently
verify if a dissimilarity space is (strictly) quasi-circular Robinson or (strictly) circular Robinson with respect to a fixed
circular order. Section \ref{s:prop} can be viewed as the preparatory work for the recognition algorithm, which is
described in Section \ref{SECTION_algo}. Using the algorithm, we prove that a strict quasi-circular Robinson space
has one or two compatible orders and their opposites and that a strict circular Robinson space has one compatible order and its opposite.
Section \ref{s:concl} provides a brief conclusion.

\section{Preliminaries} \label{s:prel} In this section, first we introduce the notions related to
the dissimilarity spaces and linear Robinson spaces. Then, we define the circular orders and the
different types of circular Robinson spaces.

\subsection{Dissimilarity spaces}

Let $X=\{ x_1,\ldots, x_n\}$ be a set of $n$ elements, called
\emph{points}. A {\it dissimilarity} on $X$ is a symmetric function
$\BD$ from $X^2$ to the nonnegative real numbers such that
$\BD(x,y)=\BD(y,x)\ge 0$ and $\BD(x,y)=0$ if and only if $x=y$. Then $\BD(x,y)$ is called
the {\it distance} between $x,y$ and $(X,\BD)$ is called a
\emph{dissimilarity space}. The \emph{ball} (respectively, the \emph{sphere}) of radius $r\ge 0$
centered at $x \in X$ is the set $B_r(x):=\{y\in X: \BD(x,y)\leq r\}$
(respectively, $S_r(x):=\{y\in X: \BD(x,y)=r\}$). The
\emph{eccentricity} of a point $x$ is $r_x:=\max \{ \BD(x,y): y\in X\}$.
Given a point $x\in X$, a point $y\in X$ is called a
\emph{farthest neighbor} of $x$ if $\BD(x,y)=r_x$. Denote by $F_x$ the
set of all farthest neighbors of $x$ and note that $F_x=S_{r_x}(x)$.
%
The \emph{distance} between two subsets $A,B$ of $X$ is
$\BD(A,B)=\min \{ \BD(a,b): a\in A, b\in B\}$.

\subsection{Compatible orders and Robinson spaces}

A partial order on $X$ is called \emph{linear} if
any two elements of $X$ are comparable. A dissimilarity $d$ and a linear
order $<$ on $X$ are called {\it compatible} if $x<y<z$ implies
$\BD(x,z)\geq \mbox{max}\{ \BD(x,y),\BD(y,z)\}.$ If $\BD$ and $<$ are
compatible, then $\BD$ is also compatible with the linear order $<^{\op}$
opposite to $<$. If a dissimilarity space $(X,\BD)$ admits a compatible
order, then $d$ is said to be {\em Robinson} and $(X,\BD)$ is called a
{\em Robinson space}. Equivalently, $(X,\BD)$ is Robinson if its
distance matrix $D$ can be symmetrically permuted so that
its elements do not decrease when moving away from the main diagonal
along any row or column. Such a dissimilarity matrix $D$ is said to
have the {\it Robinson property}. From the definition of a Robinson
dissimilarity follows that $d$ is Robinson if and only if there exists
an order $<$ on $X$ such that all balls $B_r(x)$ of $(X,\BD)$ are
intervals of $<$. Moreover, this property holds for all compatible
orders. Basic examples of Robinson dissimilarities are the
ultrametrics, thoroughly used in phylogeny. Recall, that $d$ is an
{\it ultrametric} if $\BD(x,y)\le \mbox{max}\{ \BD(x,z),\BD(y,z)\}$ for all
$x,y,z\in X$. Another example of a Robinson space is provided by the standard {\it
line-distance} between $n$ points $p_1,\ldots,p_n$ of
${\mathbb R}$ such that $p_1<\ldots <p_n$. Any line-distance has exactly two compatible orders:
the order $p_1<\ldots<p_n$ and its opposite. 

A dissimilarity $\BD$ on a set $X$ is {\em strictly Robinson} if there exits a linear order, said {\em compatible}, on $X$ such that  $x<y< z$ implies $\BD(x,z) > \max\{\BD(x,y), \BD(y,z)\}$. $(X,\BD)$ is then called a {\em strict circular Robinson space}.

\subsection{Compatible circular orders and circular Robinson spaces}

Informally speaking, a circular order on a finite set $X$ is obtained
by arranging the points of $X$ on a circle $C$. Formally, a
\emph{circular order} is a ternary relation $\beta$ on $X$ where
$\beta (u,v,w)$ expresses that the directed path from $u$ to $w$
passes through $v.$ This relation is total, asymmetric, and
transitive, which can be formulated in terms of Huntington's axioms
\cite{Hu}: for any four points $u, v, w, x$ of $X,$

\begin{enumerate}[label=(CO\arabic*)]
\item\label{it:def-co1} if $u,v,w$ are distinct, then $\beta (u,v,w)$ or $\beta (w,v,u),$
\item\label{it:def-co2} $\beta (u,v,w)$ and $\beta (w,v,u)$ is impossible,
\item\label{it:def-co3} $\beta(u,v,w)$ implies $\beta (v,w,u),$
\item\label{it:def-co4} $\beta (u,v,w )$ and $\beta (u,w,x)$ imply $\beta (u,v,x).$
\end{enumerate}

From the definition it follows that only triplets of distinct points
can be in the relation $\beta$. It also follows that the reverse relation
$\beta^{op}$, defined by setting $\beta^{op}(u,v,w)$ exactly when
$\beta (w,v,u)$, is also a circular order. Since $X$ is finite, the
circular orders on $X$ are just the orientations of the circle $C$
with points of $X$ located on $C$. We will suppose that $\beta$
corresponds to the counterclockwise order of $C$ and $\beta^{op}$ to
the clockwise order of $C$. Given a circular order $\beta$ and three
distinct points $u,v,w$, we will write $u\< v\< w$ if $\beta (u,v,w)$
holds.

For a sequence of points $x_1,x_2,\ldots,x_\ell$ containing at least
three distinct points, we will write
$x_1 \<_{\beta} x_2 \<_{\beta} \ldots \<_{\beta} x_\ell$ (or simply
$x_1 \< x_2 \< \ldots \< x_\ell$, if no ambiguity occurs) if for any
$1 \le i < j < k \le \ell$ with $x_i, x_j, x_k$ distinct,
$\beta(x_i,x_j,x_k)$ holds. We will use the following properties of circular orders:

\begin{proposition}
  Let $\beta$ be a circular order on $X$ and $x_1,x_2,\ldots, x_{\ell}\in X$ such that
  $x_1 \<_\beta x_2 \<_\beta \ldots \<_\beta x_\ell$. Then:
  \begin{enumerate}[label=(\roman*)]
  \item\label{it:prop-co1} $x_2 \< x_3 \< \ldots \< x_\ell \< x_1$,
  \item\label{it:prop-co2} if $1 \leq i < j < k \leq \ell$ and $x_i = x_j \neq x_k$ hold,
    then for any $m \in \{i,\ldots,j\}$ we have $x_m = x_i$.
  \end{enumerate}
\end{proposition}

\begin{proof}
  \cref{it:prop-co1}: We must prove that for any $1 < i < j$ with
  $x_i, x_j, x_1$ distinct, $\beta(x_i,x_j,x_1)$ holds. This follows
  from \cref{it:def-co3} and $x_1 \<_\beta x_2 \<_\beta \ldots \<_\beta x_\ell$.

  \cref{it:prop-co2}: Assume that $x_m \neq x_i = x_j$. If $x_m \neq x_k$,
  then $x_i = x_j$, $x_m$ and $x_k$ are distinct, with
  $\beta(x_i,x_m,x_k)$ and $\beta(x_m,x_j,x_k)$ (because
  $x_1 \<_\beta x_2 \<_\beta \ldots \<_\beta x_k$), by
  \cref{it:def-co3} $\beta(x_m,x_k,x_i)$ and $\beta(x_j,x_k,x_m)$,
  contradicting \cref{it:def-co2} as $x_i = x_j$. If $x_m = x_k$, by
  assumption there must be a point $x_p$ distinct from $x_i = x_j$ and
  $x_m = x_k$. By \cref{it:prop-co1}, we may assume that $p=1$. Then
  $\beta(x_1,x_i,x_k)$ and $\beta(x_1,x_m,x_j)$ hold. By \cref{it:def-co3},
  we get $\beta(x_i,x_k,x_1)$, that is $\beta(x_j,x_m,x_1)$ holds,
  contradicting \cref{it:def-co2}.
\end{proof}

We say that a nonempty proper subset $A$ of $X$ is an {\it arc} of a
circular order $\beta$ on $X$ if there are no four distinct points
$u,v \in A$ and $x,y \in X \setminus A$ such that $u \< x \< v \< y$.
From the definition, if $A$ is an arc, then so is $X \setminus A$. For two points
$x,y \in X$, consider the arcs
$\arc{xy} = \{t \in X: \beta (x,t,y)\} \cup \{x, y\}$ and
$\arc{yx} = \{t\in X: \beta (y,t,x)\} \cup \{x, y\}$. Notice that
$\arc{xy} \cup \arc{yx} = X$ and $\arc{xy} \cap \arc{yx} = \{x,y\}$.
Moreover, if $x \< y\< z$, then $\arc{xy} \subset \arc{xz}$ and
$\arc{yz} \subset \arc{xz}$. This implies that if
Notice also that if $x_1 \<_{\beta} x_2 \<_{\beta} \ldots \<_{\beta} x_m$, then $\arc{x_1x_m}$ is the union
of the arcs $\arc{x_1x_2}, \ldots, \arc{x_{m-1}x_m}$ and thus
$X$ is the union of the arcs $\arc{x_1x_2}, \ldots, \arc{x_{m-1}x_m},\arc{x_mx_1}$.
For $x,y \in X$ and $Z \subset X$, we
write $x \< y \< Z$ if for all $z \in Z$ we have $x \< y\< z$.

Arcs are for circular orders what
intervals are for linear orders. Thus the arcs can be viewed as arcs of a circle
ordered counterclockwise: the arc $\arc{xy}$ is obtained by traversing
the cycle counterclockwise from $x$ to $y$. The intersection of two
arcs is not necessarily an arc. However, we can use this geometric interpretation of arcs to prove
the following elementary observation:

\begin{lemma} \label{PROPOSITION_arc_elementary} Let $A$ and $B$
 be two arcs of a circular ordered set $(X, \beta)$. If there exists
 $x \in X\setminus (A\cup B)$, then $A\cap B$ is an arc or is empty. If there
 exists $x \in B\setminus A$, then
 $A\setminus B$ is an arc or is empty.
\end{lemma}

\begin{proof} Let $A=X^{\beta}_{a'a''}$ and $B=X^{\beta}_{b',b''}$.
  Both assertions trivially holds when $a'=a''$ or $b'=b''$. So, let
  $a'\ne a''$ and $b'\ne b''$. First let $x\notin A\cup B$. Since
  $x\notin A\cup B$, we cannot have $\beta(a',x,a'')$ or
  $\beta(b',x,b'')$. Thus, by (CO1), we can suppose that
  $\beta(x,a',a'')$ and $\beta(x,b',b'')$. Suppose also, without loss
  of generality, that $\beta(x,a',b')$. If $\beta(a',a'',b')$ holds,
  then $A$ and $B$ are disjoint. Thus, let $\beta(a',b',a'')$ holds.
  Then $B\subset A$ if $\beta(b',b'',a'')$ and $A\cap B$ is the arc
  $X^{\beta}_{b',a''}$ if $\beta(b',a'',b'')$.

  Now suppose that $x\in B\setminus A$. Since $x\notin A$, we can
  suppose that $\beta(x,a',a'')$ holds. Furthermore, we can suppose
  that (1) either $\beta (x,b'',a')$ or $\beta(x,a',b'')$ holds and
  (2) either $\beta(a'',b',x)$ or $\beta(b',a'',x)$ holds. Combining
  the subcases, we conclude that $A\setminus B$ is (a) the arc
  $A=X^{\beta}_{a'a''}$ if $\beta (x,b'',a')$ and $\beta(a'',b',x)$,
  (b) the arc $X^{\beta}_{b'',a''}$ if $\beta(x,a',b'')$ and
  $\beta(a'',b',x)$, (c) the arc $X^{\beta}_{a',b'}$ if
  $\beta (x,b'',a')$ and $\beta(b',a'',x)$, and (d) the arc
  $X^{\beta}_{b'',b'}$ if $\beta(x,a',b'')$ and $\beta(b',a'',x)$.
  This concludes the proof.
\end{proof}

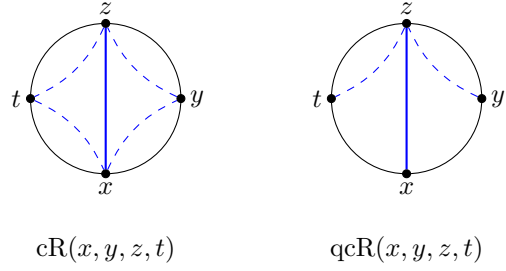
\begin{figure}
  \begin{center}
    \begin{tikzpicture}[x=1cm,y=1cm]
      \draw (0,0) circle[radius=1];
      \node[vertex] (x) at (270:1) {};
      \node[vertex] (y) at (0:1) {};
      \node[vertex] (z) at (90:1) {};
      \node[vertex] (t) at (180:1) {};
      \draw (x) node[below] {$x$};
      \draw (y) node[right] {$y$};
      \draw (z) node[above] {$z$};
      \draw (t) node[left] {$t$};
      \foreach \s/\t in {x/y,y/z,z/t,t/x} {
        \draw[dashed,blue] (\s)  to[relative,out=20,in=160] (\t);
      }
      \draw[thick,blue] (x) -- (z);
      \draw (0,-2) node { $\osR(x,y,z,t)$ };
      \begin{scope}[xshift=4cm]
        \draw (0,0) circle[radius=1];
        \node[vertex] (x) at (270:1) {};
        \node[vertex] (y) at (0:1) {};
        \node[vertex] (z) at (90:1) {};
        \node[vertex] (t) at (180:1) {};
        \draw (x) node[below] {$x$};
        \draw (y) node[right] {$y$};
        \draw (z) node[above] {$z$};
        \draw (t) node[left] {$t$};
        \foreach \s/\t in {y/z,z/t} {
          \draw[dashed,blue] (\s)  to[relative,out=20,in=160] (\t);
        }
        \draw[thick,blue] (x) -- (z);
        \draw (0,-2) node {$\qosR(x,y,z,t)$};
      \end{scope}
    \end{tikzpicture}
  \end{center}
  \caption{Illustration of the constraints defined by one-side
    Robinson and quasi one-side Robinson.}
  \label{fig:def-cR-qcR}
\end{figure}

We continue with several metric relations on four points, which will be used to
define various types of circular Robinson spaces.
Let $(X,\BD)$ be a dissimilarity space, $\beta$ be a circular order on $X$
and $x,y,z,t \in X$ such that $x\< y \< z \< t$.
\begin{itemize}[label=$\bullet$]
\item The points $x,y,z,t$ are {\em one-side Robinson}, and we denote
 it by $\osR(x,y,z,t)$, if
 $\BD(x,z) \geq \min\{\max\{\BD(x, y) , \BD(y, z) \},\max\{ \BD(x, t), \BD(t, z) \} \}$.
 See \Cref{fig:def-cR-qcR}, left.
\item The points $x,y,z,t$ are {\em strictly one-side Robinson}, and
 we denote it by $\sosR(x,y,z,t)$, if
 $\BD(x,z) > \min\{\max\{\BD(x, y) , \BD(y, z) \},\max\{ \BD(x, t), \BD(t, z) \} \}$.
\item The points $x,y,z,t$ are {\em quasi one-side Robinson}, and we
 denote it by $\qosR(x,y,z,t)$, if
 $\BD(x,z) \geq \min\{ \BD(y,z), \BD(t, z)\}$. See \Cref{fig:def-cR-qcR}, right.
\item The points $x,y,z,t$ are {\em strictly quasi one-side Robinson},
 and we denote it by $\sqosR(x,y,z,t)$, if
 $\BD(x,z) > \min\{ \BD(y,z), \BD(t, z)\}$.
\end{itemize}

Notice that the conditions $\osR(x,y,z,t)$ and $\qosR(x,y,z,t)$
trivially hold if $x=y \< z \< t$ or $x\< y \< z=t$. For
$x,y,z,t \in X$ such that $x\< y \< z \< t$, the following
implications hold:

\begin{center}
 \begin{tikzpicture}[x=1cm,y=1cm]
  \node (sosR) {$\sosR(x,y,z,t)$};
  \node (osR) [right=of sosR] {$\osR(x,y,z,t)$};
  \node (sqosR) [below=of sosR] {$\sqosR(x,y,z,t)$};
  \node (qosR) [below=of osR] {$\qosR(x,y,z,t)$};
  \draw[double equal sign distance,double,baseline,-{Implies[]}] (sosR) -- (osR);
  \draw[double equal sign distance,double,baseline,-{Implies[]}] (sosR) -- (sqosR);
  \draw[double equal sign distance,double,baseline,-{Implies[]}] (sqosR) -- (qosR);
  \draw[double equal sign distance,double,baseline,-{Implies[]}] (osR) -- (qosR);
 \end{tikzpicture}
\end{center}

Now, we define the three types of circular dissimilarities investigated in this paper and their
strict versions. A dissimilarity space $(X,\BD)$ is called {\em pre-circular Robinson} if
there exists a circular order $\beta$, which is said to be a {\em
 compatible order}, such that for all $x,y,z,t \in X$, if
$x\< y \< z \< t$ then $\osR(x,y,z,t)$ holds. The {\em strictly
 pre-circular Robinson}, {\em quasi-circular Robinson}, and {\em
 strictly quasi-circular Robinson} spaces are defined
in a similar way by using conditions $\sosR(x,y,z,t)$,
$\qosR(x,y,z,t)$, and $\sqosR(x,y,z,t)$, respectively. 
Finally, a
dissimilarity space $(X,\BD)$ is called {\em circular Robinson} (respectively {\em strictly circular Robinson}) if there
exists a circular order $\beta$, called a {\em compatible order} such
that for all $ x,y\in X$, either $(X^\beta_{xy}, \BD)$ or $(X^\beta_{yx}, \BD)$ is a
Robinson space (respectively a strict Robinson space) and the restriction of $\<$ to the arc $X^\beta_{xy}$ or
respectively $X^\beta_{yx}$ is a (linear) compatible order. 
Notice also that for all
definitions, if a circular order $\beta$ is compatible, then
$\beta^{op}$ is also compatible. A set $X$ of $n$ points
on a circle $C$ in ${\mathbb R}^2$ endowed
with the arc distance or with the chord (i.e., Euclidean) distance is a basic
example of a strict circular Robinson space.

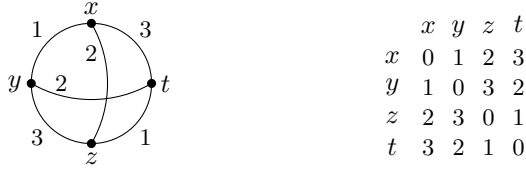
\begin{figure}[h!]
 \begin{center}
  \begin{tikzpicture}[x=0.8cm,y=0.8cm]
   \draw (0,0) circle(1) ;
   \draw (0,1) node[vertex] {} [above] node{$x$};
   \draw (0,-1) node[vertex]{} [below] node{$z$};
   \draw (1, 0) node[vertex] {} [right] node{$t$};
   \draw (-1, 0) node[vertex] {} [left] node{$y$};
   \draw (.9, .9) node{\small{3}};
   \draw (-.9, -.9) node{\small{3}};
   \draw (.9, -.9) node{\small{1}};
   \draw (-.9, .9) node{\small{1}};
   \draw (0, -1) arc(-30:30: 2);
   \draw (0, .5) node{\small{2}};
   \draw (-1,0) arc(-120: -60 : 2);
   \draw (-.5, 0) node{\small{2}};
   \begin{scope}[xshift=4cm, yshift=-.45cm]
    \draw (0, 1) node{$x$};
    \draw (0, .5) node {$y$};
    \draw (0, 0) node {$z$};
    \draw (0, -.5) node {$t$};
    \draw (.6, 1.5) node {$x$};
    \draw (1.1, 1.47) node {$y$};
    \draw (1.6, 1.5) node {$z$};
    \draw (2.1, 1.55) node {$t$};
    \draw (.6, 1) node {\small{0}};
    \draw (1.1, 1) node {\small{1}};
    \draw (1.6, 1) node {\small{2}};
    \draw (2.1, 1) node {\small{3}};
    \draw (.6, .5) node {\small{1}};
    \draw (1.1, .5) node {\small{0}};
    \draw (1.6,.5) node {\small{3}};
    \draw (2.1,.5) node {\small{2}};
    \draw (.6, 0) node {\small{2}};
    \draw (1.1, 0) node {\small{3}};
    \draw (1.6,0) node {\small{0}};
    \draw (2.1,0) node {\small{1}};
    \draw (.6,-.5) node {\small{3}};
    \draw (1.1,-.5) node {\small{2}};
    \draw (1.6,-.5) node {\small{1}};
    \draw (2.1,-.5) node {\small{0}} ;
   \end{scope}
  \end{tikzpicture}
 \end{center}
 \caption{
  A strict quasi-circular Robinson space $(X=\{x, y, z, t\},\BD)$ with
  a compatible circular order $\beta$. The quadruplet $x\<y\<z\<t$
  satisfies $\sqosR(x,y,z,t)$ but not $\sosR(x,y,z,t)$. Notice that
  the circular order obtained by reversing $z$ and $t$, i.e., such
  that $x\<y\<t\<z$, satisfies the condition $\sosR$ for all
  quadruplets.}
 \label{FIGURE_the_contre_exemple}
\end{figure}

Hubert et al. \cite{HuArMe} were the first to define circular Robinson spaces.
We do not provide their definition here because they are particular pre-circular
Robinson spaces (for their definition, see
\cite{BrOs,HuArMe}). That Robinson spaces are circular Robinson can be
seen by arranging the points of $X$ on a circle $C$ according
to a compatible order of $(X,\BD)$. Then for all $x,y\in X$, if $x<y$ in
the compatible order, then $d$ is Robinson on the arc $X^\beta_{xy}$.
As noted above, (strictly) circular Robinson spaces are (strictly) quasi-circular
Robinson spaces. However not any circular order $\beta$ satisfying
$\sqosR(x,y,z,t)$ for all quadruplets $x\< y\< z\< t$ also satisfies
$\sosR(x,y,z,t)$. Such an example is
provided in \Cref{FIGURE_the_contre_exemple}.

\section{Pre-circular Robinson spaces are circular Robinson} \label{s:circ}

In this section, we characterize pre-circular Robinson spaces. Instead
of relying directly on the condition $\osR(x,y,z,w)$, some proofs will
use the following consequence of the definition of pre-circular
Robinson spaces, stating intuitively that for any pair $u,w$, one of
the arcs $\arc{u,w}, \arc{w,u}$ has only chords shorter than $\BD(u,w)$.

\begin{lemma}\label{LEMMA_0}
  Let $(X,\BD)$ be a pre-circular Robinson space with a compatible
  circular order $\beta$ and points $u \< y \< y' \< w \< z \< z'$,
  where $u$ and $w$ are distinct. Then
  $\BD(u,w) \geq \min \{ \BD(y,y'), \BD(z,z') \}$. Moreover, if
  $(X,\BD)$ is strictly pre-circular Robinson, then this inequality is
  strict.
\end{lemma}

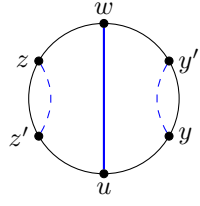
\begin{figure}
  \centering
  \begin{tikzpicture}[x=1cm,y=1cm]
    \draw (0,0) circle[radius=1];
    \node[vertex] (u) at (270:1) {};
    \node[vertex] (y) at (330:1) {};
    \node[vertex] (y2) at (30:1) {};
    \node[vertex] (w) at (90:1) {};
    \node[vertex] (z) at (150:1) {};
    \node[vertex] (z2) at (210:1) {};
    \draw (u) node[below] {$u$};
    \draw (y) node[right] {$y$};
    \draw (y2) node[right] {$y'$};
    \draw (w) node[above] {$w$};
    \draw (z) node[left] {$z$};
    \draw (z2) node[left] {$z'$};
    \foreach \s/\t in {y/y2,z/z2} {
      \draw[dashed,blue] (\s) to[relative,out=30,in=150] (\t);
    }
    \draw[thick,blue] (u) -- (w);
  \end{tikzpicture}
  \caption{Illustration of \Cref{LEMMA_0}. The solid blue
    line is longer than at least one of the dashed lines.} 
  \label{fig:LEMMA_0}
\end{figure}

\begin{proof}  We present the proof for the
  non-strict case, the strict case being slightly simpler (for an
  illustration, see \Cref{fig:LEMMA_0}). For sake of
  contradiction, assume $u \< y \< y' \< w \< z \< z'$ is a
  counterexample with a minimum number of distinct points, implying
  that $\BD(u,w) < \BD(y,y')$ and $\BD(u,w) < \BD(z,z')$. Then for
  $u\< y\< y'\< w$ we obtain the following inequalities:
 \begin{align*}
  \BD(u,y')
  &\geq \min \{ \max \{\BD(u,y),\BD(y,y')\}, \max\{\BD(y',w),\BD(w,u)\} \} \\
  &\geq \min \{ \BD(y,y'), \BD(w,u)\} \\
  &= \BD(w,u).
 \end{align*}
 The first inequality follows from $\osR(u,y,y',w)$, the second
 inequality follows from the (easily verifiable) fact that for any
 four reals $a_1,a_2,b_1,b_2$ and for any $i,j\in \{ 1,2\}$ we
 have
 $$\min \{ \max \{a_1,a_2\}, \max\{b_1,b_2\} \}\geq \min \{ a_i, b_j\}$$
 and, finally, the third inequality is implied by the initial
 condition $\BD(u,w) \geq \min \{ \BD(y,y'), \BD(z,z') \}$.
 Consequently, $\BD(u,y')\ge \BD(w,u)$. If $\BD(u,y') = \BD(w,u)$,
 then $u \< y \< w \< w \< z \< z'$ is a counterexample. If
 $\BD(u,y') > \BD(w,u)$, then $u \< u \< y' \< w \< z \< z'$ is a
 counterexample. Consequently, by the minimality of the counterexample $u \< y \< y' \< w \< z \< z'$,
 we conclude that either $u = y$ or $w = y'$ holds.
 Symmetrically, applying for $w\< z\< z'\< u$ the same reasoning as for $u\< y\< y'\< w$
 with condition
 $\osR(w,z,z',u)$ instead of  $\osR(u,y,y',w)$, we deduce that either $z = w$ or
 $z' = u$ holds. We also have that $\{y,y'\} \neq \{u,w\}$. Hence, let
 $y'' \in \{y,y'\} \setminus \{u,w\}$ and
 $z'' \in \{z,z'\} \setminus \{u,w\}$. By $\osR(u,y'',w,z'')$,
 \begin{align*}
  \BD(u,w) &\geq \min \{ \max \{\BD(u,y''),\BD(y'',w)\}, \max \{\BD(w,z''),\BD(z'',u)\} \} \\
      &\geq \min \{ \BD(y,y'), \BD(z,z') \},
 \end{align*}
 since $\{y,y'\}$ is either $\{u,y''\}$ or $\{y'',w\}$, and
 $\{z,z'\}$ is either $\{w,z''\}$ or $\{z'',u\}$. This is in
 contradiction with the assumption that $u \< y \< y' \< w \< z \< z'$
 is a counterexample. 
\end{proof}

As a consequence we have:

\begin{lemma}\label{LEMMA_pre_robinson}
 Let $(X,\BD)$ be a pre-circular Robinson space with a compatible
 circular order $\beta$ and $x,y,z$ be three arbitrary points of $X$
 such that $x \< y \< z$. If $\BD(x,z) < \max\{\BD(x,y), \BD(y,z) \}$,
 then $(\arc{zx}, \BD)$ is a Robinson space.
\end{lemma}

\begin{figure}
  \centering
    \begin{tikzpicture}[x=1cm,y=1cm,>=latex]
   \draw[nearly transparent,line width=3pt,blue] (80:1)
   arc[radius=1,start angle=80,end angle=230] (230:1);
   \draw (0,0) circle[radius=1];
   \node[vertex] (z) at (80:1) {};
   \node[vertex] (x) at (230:1) {};
   \node[vertex] (y) at (340:1) {};
   \draw (z) node[above] {$z$};
   \draw (x) node[below left] {$x$};
   \draw (y) node[right] {$y$};
   \draw[dashed,blue] (x) to[relative,out=20,in=160] (z);
   \draw[thick, blue] (x) to[relative,out=20,in=160] (y);
   \draw[thick, blue] (y) to[relative,out=20,in=160] (z);

   \path[postaction={decorate,decoration={text along path,text color=blue,text={linear Robinson}}}]
   (100:1.35) arc[radius=1.35,start angle=100,end angle=210] (210:1.35);

  \end{tikzpicture}
  \caption{Illustration of \Cref{LEMMA_pre_robinson}. When $\BD(x,z)$
    is not the maximum of the three dissimilarities, then the arc
    $\arc{zx}$ is a linear Robinson space.}
  \label{fig:LEMMA_pre_robinson}
\end{figure}
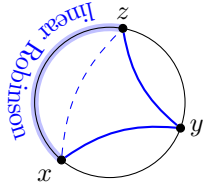

\begin{proof}
  Let $y', y'' \in \{x,y,z\}$ be
  such that $\BD(x,z) < \BD(y',y'')$ and $x \< y' \< y'' \< z$ (for an illustration,
  see \Cref{fig:LEMMA_pre_robinson}). Pick
  any points $u,v,w \in \arc{zx}$ such that $z \< u \< v \< w \< x$
  (we may have $u=z$ or $w=x$) and suppose by way of contradiction
  that $\BD(u,w) < \max\{ \BD(u,v), \BD(v,w)\}$, let
  $v',v'' \in \{u,v,w\}$ be such that $u \< v' \< v'' \< w$ and
  $\BD(u,w) < \BD(v',v'')$. If $\BD(u,w) \leq \BD(x,z)$, then
  $\BD(u,w) < \BD(y',y'')$, and by \Cref{LEMMA_0} on
  $w \< y' \< y'' \< u \< v' \< v''$, this is a contradiction. If
  $\BD(u,w) > \BD(x,z)$, then $\BD(x,z) < \BD(v',v'')$, and by
  \Cref{LEMMA_0} on $x \< y' \< y'' \< z \< v' \< v''$ we obtain again
  a contradiction.
\end{proof}

Now, we can prove our first main result:

\begin{theorem}\label{PROPOSITION_pre_robinson}
 A dissimilarity space $(X,\BD)$ is pre-circular Robinson if and only
 if $(X,\BD)$ is circular Robinson.
\end{theorem}

\begin{proof}
 To prove the theorem, first suppose that $(X,\BD)$ is a circular
 Robinson space and $\beta$ is a compatible circular order on $X$. Pick
 any $x,y,z,t \in X$ such that $x\< y \< z \< t$. By definition of
 $\beta$, $\<$ is a compatible linear order on the arc $\arc{xz}$ or
 $\arc{zx}$. In the first case, since $y\in \arc{xz}$, we have
 $\BD(x,z)\ge \max\{ \BD(x,y),\BD(y,z)\}$. In the second case, since
 $t\in \arc{zx}$, we have $\BD(x,z)\ge \max\{ \BD(x,t),\BD(t,z)\}$.
 Consequently, $\BD(x,z) \geq \min\{\max\{\BD(x,y) , \BD(y,z) \},\max\{ \BD(x,t), \BD(t,z) \} \}$,
 establishing that $(X,\BD)$ is a pre-circular Robinson space.

 Conversely, let $(X,\BD)$ be a pre-circular Robinson space and $\beta$
 be a compatible circular order. Pick any pair of points $a,b$ of
 $X$. If $(\arc{ab},\BD)$ is not a Robinson space, then there exists
 three points $x,y,z\in \arc{ab}$ such that $x\< y\< z$ and
 $\BD(x,z)<\max \{ \BD(x,y), \BD(y,z)\}$. By \Cref{LEMMA_pre_robinson},
 $(\arc{zx},\BD)$ is a Robinson space. Since $\arc{ba}\subset \arc{zx}$, this
 proves that $(\arc{ba},\BD)$ is Robinson, establishing that $(X,\BD)$ is a
 circular Robinson space and $\beta$ is a compatible circular order.
\end{proof}

As a consequence, 
$(X,\BD)$ is a \emph{strictly circular
 Robinson space} if 
 and only if 
 it is a strictly pre-circular Robinson space.

\section{Properties of quasi-circular and strict circular Robinson spaces} \label{s:prop}

In this section, we present several properties of (strict)
quasi-circular and circular Robinson spaces. We also show how to
verify if a dissimilarity space is (strictly) quasi-circular Robinson
or (strictly) circular Robinson with respect to a given circular order.

\subsection{Properties of (strictly) quasi-circular Robinson spaces}

In this subsection, we recall or present some properties of (strictly)
quasi-circular Robinson spaces. Notice that these properties are also
true for (strictly) circular Robinson spaces.
We start with the following characterization of quasi-circular
Robinson spaces of~\cite{BrOs}:

\begin{proposition} \cite{BrOs} \label{PROPOSITION_ball}
 A dissimilarity space $(X,\BD)$ is quasi-circular Robinson if
 and only if there exists a circular order 
 $\beta$ such that for
 any $x\in X$ and $r \in \mathbb{R}^+$, the ball $B_r(x)$ and its
 complement $X\setminus B_r(x)=\{t \in X: \BD(x,t)>r\}$ are arcs of
 $\beta$.
\end{proposition}

\begin{proof}
First suppose that $(X,\BD)$ is a quasi-circular Robinson space and $\beta$ is a compatible circular order. Let $B_r(x)$ be any ball of $(X,\BD)$.
We will show that $X\setminus B_r(x)$ is an arc; since the complement of an arc is an arc, this will also show that  $B_r(x)$ is an arc.
Let $y, y' \in X\setminus B_r(x)$ and suppose, with no loss of generality, that $x\< y\< y'$. Let $z \in X^\beta_{yy'}$. If $z \notin X\setminus B_r(x)$, then
$\BD(z, x)\le r < \BD(y,x), \BD(y', x)$, contradicting the condition $\qosR(z, y', x, y)$. Consequently, $X\setminus B_r(x)$ and $B_r(x)$ are arcs.
Conversely, suppose that there exists a circular order $\beta$ such
 that each ball $B_r(x)$ is an arc of $\beta$. Pick arbitrary points
 $x,y,z,t\in X$ such that $x\< y\< z\< t$. Let $r=\BD(x,z)$. Since
 $B_r(z)$ is an arc of $\beta$ and $\beta(x,y,z), \beta(z,t,x)$ hold,
 either $y$ or $t$ must belong to the ball $B_r(z)$. Consequently,
 $\BD(x,z) \geq \min\{ \BD(y,z), \BD(t, z)\}$, establishing
 $\qosR(x,y,z,t)$.
\end{proof}

Let $(X,\BD)$ be a quasi-circular Robinson space and $\beta$ be a
compatible circular order. For any point $x\in X$, recall that $F_x$
consists of all farthest neighbors of $x$ and $r_x$ is the
eccentricity of $x$. Let $M_x:=X\setminus(F_x\cup\{x\})$.
Note that $M_x\cup \{ x\}$ is a ball $B_r(x)$ for some $r$ that is strictly smaller
but sufficiently close to $r_x$.  Thus, by \cref{PROPOSITION_ball},
$M_x\cup \{ x\}$ and $F_x$ are complementary arcs of $\beta$.
Consequently, the set $M_x$ is partitioned into two arcs
$L_x :=\{t \in M_x : x\< t \< F_x \}$ and
$R_x :=\{t \in M_x : F_x \< t \< x\}$ (left and right arcs), where one
of those arcs may be empty. Two points $y,y' \in M_x$ are called {\em
 $x$-separated} if they belong to distinct arcs $L_x$ and $R_x$.

The algorithmic importance of the sets $L_x$ and $R_x$ is due to the
fact that, as the following lemma shows, the circular order of each of
these sets is given by the order of the distances to $x$.

\begin{lemma}\label{PROPOSITION_order}
 Let $(X,\BD)$ be a quasi-circular Robinson space, $\beta$ be a
 compatible circular order, and $x$ any point of $X$. If
 $y,z \in L_x$ and $x\< y\<z$ or $y,z \in R_x$ and $z\< y\<x$, then
 $\BD(x,y) \leq \BD(x,z)$. Moreover, if $(X,\BD)$ is strict quasi-circular,
 then $\BD(x,y) < \BD(x,z)$.
\end{lemma}

\begin{proof}
 Let $y,z \in L_x$ with $x\< y\<z$. Let $t \in F_x$. By
 $\qosR(x,y,z,t)$ and since $\BD(x,z) < \BD(x,t)$, we obtain that $\BD(x,y) \leq \BD(x,z)$.
 The proof for $y, z \in R_x$ is similar.
\end{proof}

\begin{lemma}\label{PROPOSITION_opposite}
 Let $(X, \BD)$ be a strict quasi-circular Robinson space, $\beta$ be a
 compatible circular order, and $x$ any point of $X$. Then any sphere
 $S_r(x)$ contains at most two points. Furthermore, if $r<r_x$ and
 $S_r(x)$ consists of two points $y,y'$, then $y$ and $y'$ are
 $x$-separated.
\end{lemma}

\begin{proof} Suppose by way of contradiction that there exist three points
$y, y', y''\in S_r(x)$. We can suppose, with no loss of generality, that $y\< y'\< y''$.
Since $X$ is covered by the arcs $\arc{yy'},\arc{y'y''},$ and $\arc{y''y}$, we can suppose that
$x$ belongs to $\arc{y''y}$, i.e., that $x \< y \< y'\< y''$. By condition
$\sqosR(y', y'', x, y)$, we must have $d(x,y')>\min \{ d(x,y''),d(x,y)\}$.
Since $d(x,y)=d(x,y')=d(x,y'')=r$, this is impossible.  Thus $|S_r(x)|\le 2$.
Since $(X,\BD)$ is strict
quasi-circular, by \Cref{PROPOSITION_order},
$|S_r(x) \cap L_x| \leq 1$ and $|S_r(x) \cap R_x| \leq 1$, proving the
second assertion.
\end{proof}

\subsection{Differences between  (strictly) circular and quasi-circular Robinson spaces}

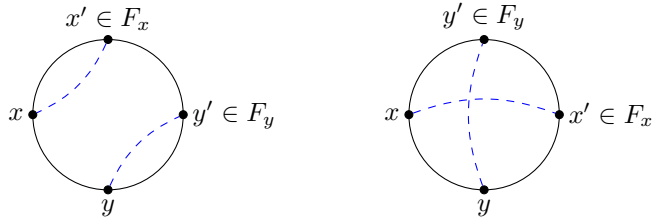
\begin{figure}[h]
  \begin{center}
    \begin{tikzpicture}[x=1cm,y=1cm]
      \draw (0,0) circle[radius=1];
      \node[vertex] (x) at (270:1) {};
      \node[vertex] (y) at (0:1) {};
      \node[vertex] (z) at (90:1) {};
      \node[vertex] (t) at (180:1) {};
      \draw (x) node[below] {$y$};
      \draw (y) node[right] {$y' \in F_y$};
      \draw (z) node[above] {$x' \in F_x$};
      \draw (t) node[left] {$x$};
       \draw[dashed,blue] (x)  to[relative,out=20,in=160] (y);
       \draw[dashed,blue] (z)  to[relative,out=20,in=160] (t);
       \begin{scope}[xshift= 5cm]
         \draw (0,0) circle[radius=1];
      	 \node[vertex] (x) at (270:1) {};
     	 \node[vertex] (y) at (0:1) {};
     	 \node[vertex] (z) at (90:1) {};
     	 \node[vertex] (t) at (180:1) {};
     	 \draw (x) node[below] {$y$};
     	 \draw (y) node[right] {$x' \in F_x$};
     	 \draw (z) node[above] {$y' \in F_y$};
     	 \draw (t) node[left] {$x$};
     	  \draw[dashed,blue] (x)  to[relative,out=20,in=160] (z);
     	  \draw[dashed,blue] (t)  to[relative,out=20,in=160] (y);
       \end{scope}
    \end{tikzpicture}
  \end{center}
  \caption{Illustration of Propositions~\ref{PROPOSITION_noncrossingdiags}. and \ref{PROPOSITION_crossingdiags}. On the left, since $xx'$ and $yy'$ do not cross, the space can be quasi-circular but not circular. On the right, the space can be circular.}
  \label{fig:quasi_pas_circulaire}
\end{figure}

In this subsection, we present two properties which distinguish (strictly)
circular Robinson spaces and (strictly) quasi-circular Robinson spaces.
Roughly speaking, in a circular space, for $x,y \in X$, $x'\in F_x$ and $y'\in F_y$, $xx'$ and $yy'$ have to cross each other, but this is not the case for  quasi-circular spaces (see \Cref{fig:quasi_pas_circulaire}).

\begin{proposition}\label{PROPOSITION_noncrossingdiags}
 Let $(X,\BD)$ be a circular Robinson space with a compatible order
 $\beta$. Then for all $x, y \in X$, $x' \in F_x$, $y' \in F_y$ with
 $|\{x,x',y,y'\}| \geq 3$, one of the following assertions holds:
 \begin{enumerate}[label=$(\alph*)$,ref=$(\alph*)$,nosep]
 \item\label{it:a} $x \< y \< x' \< y'$,
 \item\label{it:b} $x \< y' \< x' \< y$,
 \item\label{it:c} $\{ y,y'\}\cap F_x\ne \varnothing$ or $\{ x,x'\}\cap F_y\ne \varnothing$.
 \end{enumerate}
 Moreover if $(X,\BD)$ is strictly circular Robinson, then either
 \cref{it:a} or \cref{it:b} holds.
\end{proposition}

\begin{proof}
 Suppose first that $(X,\BD)$ is strictly circular Robinson. For sake
 of contradiction, assume none of these assertions holds. There are
 two cases depending on the order (up to reversal) of $x,y,x',y'$.

 If $x \< y \< y' \< x'$, by $\sosR(x,y,y',x')$
 and $\sosR(y,y',x',x)$, and since $x' \in F_x$,
 $y' \in F_y$, we get:
 \begin{align*}
  \BD(x,x') &\geq \BD(x,y') \\
      &> \min \{ \max\{\BD(x,y),\BD(y,y')\}, \max\{\BD(x,x'),\BD(x',y')\}\} \\
      &\geq \min \{ \BD(y,y'), \max \{\BD(x,x'),\BD(x',y')\} \} \\
      &\geq \min \{ \BD(y,y'), \BD(x,x') \},\\
  \BD(y,y') &\geq \BD(y,x') \\
      &> \min \{ \max\{\BD(y,y'),\BD(y',x')\}, \max\{\BD(y,x),\BD(x,x')\}\}\\
      &\geq \min \{ \max \{\BD(y,y'\}, \BD(y',x')\}, \BD(x,x')\}\\
      &\geq \min \{ \BD(y,y'), \BD(x,x') \}.
 \end{align*}
 From this, we get that $\BD(x,x') > \BD(y,y')$ and $\BD(y,y') > \BD(x,x')$, a contradiction.

 If $x \< y' \< y \< x'$, then by $\sosR(x,y',y,x')$ and
 using that $x' \in F_x, y' \in F_y$, we get:
 \begin{align*}
  \min \{\BD(x,x'), \BD(y,y')\}
  &\geq \BD(x,y) \\
  &> \min \{ \max\{\BD(x,y'),\BD(y',y)\}, \max\{\BD(x,x'),\BD(x',y)\}\} \\
  &\geq \min \{ \BD(y',y), \BD(x,x') \},
 \end{align*}
 a contradiction.

 Suppose now that $(X,\BD)$ is non-strictly circular Robinson. We
 follow the same argument. In the first case, instead of a
 contradiction, we get that $\BD(x,x') = \BD(y,y')$. Applying $\osR(x,y,y',x')$
 and $\osR(y,y',x',x)$, we conclude that
 $y' \in F_x$ and $x' \in F_y$, 
 implying \Cref{it:c}. In the second case, we get that
 $\BD(x,y) = \min \{\BD(x,x'), \BD(y,y')\}$, implying either $y \in F_x$ or
 $x \in F_y$, that is \Cref{it:c}.
\end{proof}

Now, let $\beta$ be a circular order on $X$ that is compatible with a
quasi-circular Robinson space $(X,\BD)$. We determine under which
conditions $\beta$ is not compatible with respect to the circular
Robinson property of $(X,\BD)$.

\begin{proposition}\label{PROPOSITION_crossingdiags}
 Let $(X,\BD)$ be a (strict) quasi-circular Robinson space and $\beta$
 a compatible order, such that $(X,\BD)$ is not (strict) circular
 Robinson with respect to $\beta$. Then there exist $x, y \in X$,
 $x' \in F_x$, $y' \in F_y$ such that $x \< x' \< y \< y'$ or
 $x \< y' \< y \< x'$. Moreover, in the non-strict case,
 we may also assume that $x,x' \notin F_y$ and $y,y' \notin F_x$.
\end{proposition}

\begin{proof}
 We first prove the strict case. Let $x \< u \< y \< v$ be such that
 $\sosR(x,u,y,v)$ does not hold:
 \begin{equation}
  \BD(x,y) \leq \min \{ \max \{\BD(x,u),\BD(u,y)\}, \max\{\BD(x,v),\BD(v,y)\} \}. \label[ineq]{eq:nonstrict}
 \end{equation}
 By $\sqosR(x,u,y,v)$ and $\sqosR(y,v,x,u)$, we get:
 \begin{align}
  \BD(x,y) &> \min \{ \BD(x,u), \BD(x,v) \}, \label[ineq]{eq:strict1}\\
  \BD(x,y) &> \min \{ \BD(y,u),\BD(y,v) \}. \label[ineq]{eq:strict2}
 \end{align}
 Combining these inequalities, we get:
 \begin{align*}
  \min \{\BD(x,u),\BD(x,v)\} < \max \{\BD(x,u),\BD(u,y)\}, \qquad &
  \min \{\BD(x,u),\BD(x,v)\} < \max \{\BD(x,v),\BD(v,y)\}, \\
  \min \{\BD(y,u),\BD(y,v)\} < \max \{\BD(x,u),\BD(u,y)\}, \qquad &
  \min \{\BD(y,u),\BD(y,v)\} < \max \{\BD(x,v),\BD(v,y)\},
 \end{align*}
 and then:
 \begin{align*}
  \BD(x,v) < \BD(x,u) \lor \BD(x,u) < \BD(u,y), \qquad &
  \BD(x,u) < \BD(x,v) \lor \BD(x,v) < \BD(v,y), \\
  \BD(y,v) < \BD(y,u) \lor \BD(y,u) < \BD(x,u), \qquad &
  \BD(y,u) < \BD(y,v) \lor \BD(y,v) < \BD(x,v),
 \end{align*}
 which is equivalent to the disjunction of these two symmetric assertions:
 \begin{enumerate}[label=(\roman*),nosep]
 \item $\BD(x,v) < \BD(x,u)$, $\BD(x,v) < \BD(v,y)$, $\BD(y,u) < \BD(y,v)$,
  and $\BD(y,u) < \BD(x,u)$, 
 \item $\BD(x,v) > \BD(x,u)$, $\BD(x,v) > \BD(v,y)$,
  $\BD(y,u) > \BD(y,v)$, and $\BD(y,u) > \BD(x,u)$. 
 \end{enumerate}
 We may assume the first. Then
 $\BD(x,v) = \min \{\BD(x,v), \BD(x,u)\} < \BD(x,y) \leq \min \{\BD(x,u), \BD(v,y) \} \leq \BD(x,u)$
 (by \Cref{eq:nonstrict,eq:strict1}), which implies by the strict
 unimodality of distances from $x$ that $F_x \subseteq X^\beta_{uy}$.
 Similarly, $\BD(y,u) < \BD(x,y) \leq \BD(y,v)$ which implies that
 $F_y \subseteq X^\beta_{vx}$. Consequently, if $x'\in F_x$ and
 $y'\in F_y$, then we get $x\< x'\< y\<y'$, as expected.

 In the non-strict case, \Cref{eq:nonstrict} becomes a strict
 inequality, while \Cref{eq:strict1,eq:strict2} become non-strict
 inequalities. Combining these inequalities, we get the same
 conclusion as in the strict case. For example, in the first case we
 get that
 $\BD(x,v) = \min \{\BD(x,v), \BD(x,u)\} \leq \BD(x,y) < \min \{\BD(x,u), \BD(v,y) \} \leq \BD(x,u)$,
 yielding $\BD(x,v) \leq \BD(x,y) < \BD(x,u)$ and
 $\BD(y,u) \leq \BD(x,y) < \BD(y,v)$. By unimodality of distances, we
 conclude that $F_x \subseteq X^\beta_{uy}$ and
 $F_y \subseteq X^\beta_{vx}$. Consequently, if $x'\in F_x$ and
 $y'\in F_y$, then we get $x\< x'\< y\<y'$. Moreover, $y \notin F_x$
 and $x \notin F_y$. Since $x\< x'\< v\< y'$ and $\BD(x,v)<\BD(x,x')$, by
 $\qosR(x,x',v,y')$ we conclude that $\BD(x,v)\ge \BD(x,y')$, yielding
 $y'\notin F_x$. Analogously, one can show that $x'\notin F_y$.
\end{proof}

\subsection{Verification of compatibility}\label{verification}

In this subsection, given a circular order $\beta$, we describe how to
check in $O(n^2)$ whether a dissimilarity space $(X,\BD)$ on $n$
points is (strictly) quasi-circular Robinson or (strictly) circular
Robinson with respect to $\beta$. This verification task can also be
done in $O(n^2)$ for strict circular Robinson spaces, as defined in
\cite{HuArMe}.
Then, we will show how to extend this result to strict versions of the
other definitions of circular dissimilarities introduced by Brucker
and Osswald~\cite{BrOs}, namely the dissimilarities whose 2-balls or
clusters are arcs.

To test whether $(X,\BD)$ is (strictly) quasi-circular Robinson with
respect to $\beta$, by \Cref{PROPOSITION_ball} we have to test whether
all balls of $(X,\BD)$ are arcs of $\beta$. This can be done in the
following way. Let $D$ be the distance matrix of $(X,\BD)$ ordered
according to the circular order $\beta$. The matrix $D$ is called
\emph{unimodal} if for each row $i$, when moving circularly from the
element $d_{ii}$ on the main diagonal of $D$ to the right until the
last element $d_{in}$ and then from the first element $d_{i1}$ until
$d_{ii}$, the elements first increase monotonically, stay at the
maximal values, and then decrease monotonically. Since $D$ is
symmetric, the same monotonicity property holds also for each column
$i$: moving down from $d_{ii}$ until $d_{ni}$ and then from $d_{1i}$
until $d_{ii}$, the elements first increase monotonically, stay at the
maximal value, and then decrease monotonically. We say that $D$ is
{\em strictly unimodal} if the values strictly increase, have one or
two maximal elements, and then strictly decrease. It was shown in
\cite[Proposition 3.7]{ArGuSiLo} that $\beta$ is a compatible circular
order for a quasi-circular Robinson space (respectively, strictly
quasi-circular Robinson space) if and only if $D$ is unimodal
(respectively, strictly unimodal). From the definition, testing if $D$
is (strictly) unimodal can be easily done in $O(n^2)$ time. In case of
strict unimodality we also have to check that each row has at most two
maximal elements (this correspond to computing for each $x\in X$ the
set $F_x$ and checking if $|F_x|\le 2$). Notice that for strictly
circular Robinson spaces defined in \cite{HuArMe}, this testing task
can be also done in $O(n^2)$ time.

Next, we consider the task of testing whether $(X,\BD)$ is (strictly)
circular Robinson with respect to a circular order $\beta$. 
As (strictly) circular Robinson spaces are particular cases of
(strictly) quasi-Robinson spaces, the (strict) unimodality of the
distance matrix $D$ is a necessary condition for compatibility. Under
this condition, we can use \Cref{PROPOSITION_crossingdiags}. Namely,
we compute the arc $F_x$ for each $x \in X$, and store the indices of
its extremities. This can be done by dichotomy (using $\beta$) in
$O(n \log n)$ total time. Then for each pair $x, y \in X$, we can
check in constant time whether there are $x' \in F_x$, $y' \in F_y$ as
given in \Cref{PROPOSITION_crossingdiags}. If such elements exist,
then by \Cref{PROPOSITION_noncrossingdiags}, $(X,\BD)$ is not
(strictly) circular Robinson.
Otherwise, 
$(X,\BD)$ is (strictly) circular Robinson with respect to $\beta$.
This testing task can be done in $O(n^2)$ time. As a consequence, we
have the following result:

\begin{proposition}\label{testing}
 For a dissimilarity space $(X,\BD)$ on $n$ points and a circular order
 $\beta$ on $X$, one can check in $O(n^2)$ time whether, with respect
 to $\beta$, $(X,\BD)$ is (1) (strictly) quasi-circular Robinson, (2)
 (strictly) circular Robinson.
\end{proposition}

\section{The recognition algorithm}\label{SECTION_algo}

In this section, we describe a simple but optimal algorithm to
recognize strictly circular Robinson spaces and strictly
quasi-circular Robinson spaces. Our algorithm consists in partitioning
$X$ into four sets with respect to any point $x\in X$ and any $x'\in F_x$. We
prove that those four sets are arcs in any compatible circular order
$\beta$ and that the restriction of $\beta$ to each of these four sets
is obtained by sorting its points by distances to $x$. Concatenating
these four arcs, we obtain two circular orders. Finally, it suffices
to verify that one of these circular orders is compatible. This also
shows that any strict circular Robinson space (in each of the three
versions) has one or two compatible circular orders and their
opposites.

\subsection{How to define arcs \texorpdfstring{$\arc{xy}$}{} metrically}

Given a dissimilarity space $(X,\BD)$ and two distinct points
$x ,y \in X$, we set
$\I(x,y) = \{ u \in X : \BD(x,y) > \max \{\BD(x,u), \BD(u,y)\}\}$ and
$\J(x,y) = \I(x,y) \cup \{x,y\}$. In all results of this subsection,
we assume that $(X,\BD)$ is a strict quasi-circular Robinson space and
$\beta$ is an arbitrary compatible circular order on $X$.

\begin{lemma}\label{LEMMA_3Points}
  Let $x \< y \< z$ be three points of
 $X$ such that $\BD(x,y) \leq \min \{ \BD(x,z), \BD(y,z) \}$. Then
 $\arc{xy}=\J(x,y)$.
\end{lemma}

\begin{proof}
 First, let $v \in \arc{xy} \setminus \{x,y\}$, i.e., $x \< v \< y$.
 By $\sqosR(x,v,y,z)$ and since $\BD(x,y) \leq \BD(x,z)$, we have
 $\BD(x,v) < \BD(x,y)$. By $\sqosR(y,z,v,x)$ and since
 $\BD(y,x) \leq \BD(y,z)$, we have $\BD(y,u) < \BD(y,x)$. Hence
 $\arc{xy} \subseteq \J(x,y)$. To prove the converse inclusion, let
 $u \in \arc{yz} \setminus \{y,z\}$, that is $y \< u \< z$. By
 $\sqosR(u,z,x,y)$, $\BD(x,u) > \min \{\BD(x,y), \BD(x,z)\} = \BD(x,y)$,
 hence $u \notin \I(x,y)$. Similarly if
 $u \in \arc{zx}\setminus \{z,x\}$, applying $\sqosR(y,z,u,x)$ we also
 get $u \notin \I(x,y)$. Consequently,
 $(\arc{yz} \cup \arc{zx}) \cap \I(x,y) = \{x,y\}$, establishing the
 inclusion $\J(x,y)\subseteq \arc{xy}$. Thus $\arc{xy} = \J(x,y)$.
\end{proof}

Now, let $x$ be an arbitrary point of $X$ and $x' \in F_x$. Let
$N = \{ u \in X : \BD(u,x) \leq \BD(u,x') \}$ and
$F = \{ u \in X : \BD(u,x) \geq \BD(u,x') \}$. Note that $N \cup F = X$
and $x\in N\setminus F, x'\in F\setminus N$.

\begin{lemma}\label{CLAIM_NFarcs}
$N$ and $F$ are arcs of $\beta$.
\end{lemma}

\begin{proof}
  It suffices to prove that $N$ is an arc, as $F = X \setminus N$ and
  $x' \in F \neq \varnothing$. Let $y,z \in X \setminus \{x,x'\}$ be
  distinct points with $x \< y \< z \< x'$ and $z\in N$. We assert
  that $y \in N$. Since $z\< x'\< x\< y$ by (CO3) and
  $\BD(x,x')\ge \BD(x,y)$ because $x'\in F_x$, by $\sqosR(z,x',x,y)$,
  we have $\BD(x,y) < \BD(x,z)$. Since we also have $y\< z\< x'\< x$
  by (CO3), by condition $\sqosR(y,z,x',x)$,
 $$\BD(x',y) > \min \{ \BD(x',z), \BD(x',x) \} \geq \min \{\BD(x,z),\BD(x,z)\} = \BD(x,z) > \BD(x,y).$$
 The second inequality follows from the fact that $d(x',z)\ge d(x,z)$
 since $z \in N$ and $\BD(x',x)\ge \BD(x,z)$ since $x' \in F_x$.
 Consequently, $\BD(x',y)>\BD(x,y)$, implying that $y \in N$.
 Symmetrically, if $z \< y \< x$ with $z \in N$, then $y \in N$.
 Hence, $N$ is an arc of $\beta$.
\end{proof}

\begin{figure}[htbp]
 \begin{center}
   \begin{tabular}{cp{0.1\textwidth}c}
    \begin{tikzpicture}[x=1cm,y=1cm]
     \draw[nearly transparent,line width=3pt,blue] (90:1)
      arc[radius=1,start angle=90,end angle=180] (180:1);
     \draw[nearly transparent,line width=3pt,red] (180:1)
      arc[radius=1,start angle=180,end angle=270] (270:1);
     \draw (0,0) circle[radius=1];
     \tkzDefPoint(90:1){x}
     \tkzDefPoint(180:1){y}
     \tkzDefPoint(270:1){x1}
     \tkzDrawPoints[vertex](x,y,x1)
     \node[above=0pt of x] {$x$};
     \node[left=0pt of y] {$y$};
     \node[below=0pt of x1] {$x'$};
     \draw (135:1.6) node[text=blue] {$J(x,y)$};
     \draw (225:1.6) node[text=red] {$J(y,x')$};
     \tkzDrawSegments(x,y y,x1)
     \tkzMarkSegments(x,y y,x1)
    \end{tikzpicture}
    & &
      \begin{tikzpicture}[x=1cm,y=1cm]
       \draw[nearly transparent,line width=3pt,blue] (10:1)
        arc[radius=1,start angle=10,end angle=90] (90:1);
       \draw[nearly transparent,line width=3pt,red] (190:1)
        arc[radius=1,start angle=190,end angle=270] (270:1);
       \draw (0,0) circle[radius=1];
       \draw[gray,thin] (-1.3,0) -- (1.3,0);
       \tkzDefPoint(90:1){x}
       \tkzDefPoint(170:1){z1}
       \tkzDefPoint(190:1){y}
       \tkzDefPoint(270:1){x1}
       \tkzDefPoint(350:1){y1}
       \tkzDefPoint(10:1){z}
       \tkzDrawPoints[vertex](x, y, z, x1, z1, y1)
       \tkzDrawSegments(x,z x1,y)
       \node[above=0pt of x] {$x$};
       \node[right=0pt of z] {$z$};
       \node[left=0pt of y] {$y$};
       \node[below=0pt of x1] {$x'$};
       \node[left=0pt of z1] {$z'$};
       \node[right=0pt of y1] {$y'$};
       \draw (45:1.6) node[text=blue] {$J(x,z)$};
       \draw (225:1.6) node[text=red] {$J(y,x')$};
       \draw (90:0.5) node {$N$};
       \draw (270:0.5) node {$F$};
      \end{tikzpicture}
    \\
    (a) & & (b)
   \end{tabular}
  \end{center}
  \caption{Configurations occurring in (a) \Cref{CLAIM_equalcase} and (b)
   \Cref{LEMMA_fourpartition}. In (b), the positions of $z$ and
   $z'$ may be swapped, as well as those of $y$ and $y'$.}
 \label{FIG_partition_into_arcs}
\end{figure}
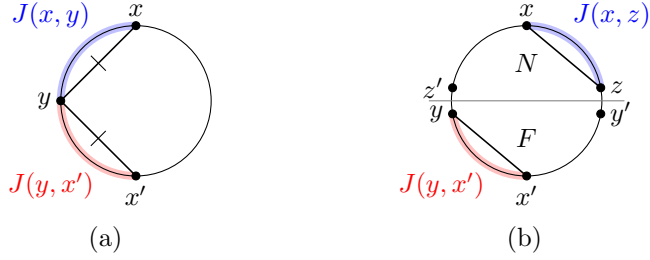

\begin{lemma}\label{CLAIM_equalcase}
  If $N \cap F\ne \varnothing$ and $y$ is a point of $X$ with
  $\BD(x,y) = \BD(y,x')$, then $J(x,y) \cup J(y,x')$ either
  coincides with $\arc{xx'}$ when $x \< y \< x'$ or with $\arc{x'x}$
  when $x' \< y \< x$.
\end{lemma}

\begin{proof}
 Suppose without loss of generality that $x \< y \< x'$ (see
 \Cref{FIG_partition_into_arcs} (a)). Since $\BD(x,y)=\BD(y,x')$ and
 $\BD(x,y)\le \BD(x,x')$, by \Cref{LEMMA_3Points} we conclude that
 $\arc{xy}=\J(x,y)$. By (CO3), we have $y\< x'\< x$. From the choice
 of the points $x'\in F_x$ and $y$ we have
 $\BD(y,x')\le \min \{ \BD(x',x),\BD(y,x)\}$. By \Cref{LEMMA_3Points} we
 conclude that $\arc{yx'} = \J(y,x')$. Finally, since $x \< y \< x'$,
 we have $\arc{xx'}=\arc{xy}\cup \arc{yx'}$, yielding
 $\arc{xx'}=J(x,y) \cup J(y,x')$.
\end{proof}

Consequently, if $N \cap F \ne \varnothing$, and $y$ is a point with
$\BD(x,y) = \BD(y,x')$, then according to \Cref{CLAIM_equalcase}, the
circular order $\beta$ such that $x \< y \< x'$ can be computed in
$O(n \log n)$ time. This is done by computing
$\arc{xx'} = J(x,y) \cup J(y,x')$, then ordering the points of
$\arc{xx'}$ and of its complement $X\setminus \arc{xx'}$ by distances
to $x$, by \Cref{PROPOSITION_order}. Notice that in this case the
compatible circular order $\beta$ is unique up to reversal.

Thus, we may next assume that $N \cap F = \varnothing$. The points
$w \in N$ such that $x \< w \< x'$ form an arc whose ordering is given
by increasing distances from $x$. Analogously, the points $w \in N$
such that $x' \< w \< x$ form an arc whose ordering is given by
decreasing distances from $x$. The points of $F$ are similarly
distributed into two arcs with respect to the distances from $x'$.
Therefore, it is sufficient to partition the sets $N \setminus \{x\}$
and $F \setminus \{x'\}$ into such pairs $N', N''$ and $F',F''$, respectively.
This is done by the next lemma. 

\begin{lemma}\label{LEMMA_fourpartition}
  If $N \cap F = \varnothing$, then there exist $z, z' \in N$ and
  $y, y' \in F$ and two bipartitions $N\setminus \{x\} = N' \cup N''$,
  $F\setminus \{x'\} = F' \cup F''$ such that for any compatible order $\beta$ on $X$
  we have $\{N'\cup\{ x\}, N''\cup\{ x\}\} = \{ \arc{zx}, \arc{xz'}\}$ and
  $\{F'\cup \{ x'\}, F''\cup \{ x'\}\} = \{ \arc{yx'}, \arc{x'y'}\}$. The sets
  $N', N'', F', F''$ and the points $z, z', y, y'$ can be computed in
  $O(n)$ time.
\end{lemma}


\begin{proof}
 Assume that $N \neq \{ x \}$, and let $z \in N$ with $\BD(x,z)$
 maximal (see \Cref{FIG_partition_into_arcs} (b)). Then applying
 \Cref{LEMMA_3Points} to $x,z,x'$, we have that $\J(x,z)$ is either
 $\arc{xz}$ or $\arc{zx}$ (depending of whether $x \< z \< x'$ or
 $x' \< z \< x$). We denote $N' = \J(x,z)$. If
 $N''=N \setminus N' \neq \varnothing$, let $z' \in N''$ with
 $\BD(x,z')$ maximal. By \Cref{LEMMA_3Points}, $\J(x,z')$ is either
 $\arc{xz'}$ or $\arc{z'x}$. By \Cref{PROPOSITION_order}, $z$ and
 $z'$ are $x$-separated, that is:
 \begin{itemize}[label=-]
 \item[$\bullet$] either $\J(x,z') = \arc{z'x}$ and $\J(x,z) = \arc{xz}$,
 \item[$\bullet$] or $\J(x,z) = \arc{zx}$ and $\J(x,z') = \arc{xz'}$.
 \end{itemize}
 By the choice of $z$ and $z'$, we conclude that $N = \J(x,z) \cup \J(x,z')$.
 If $z$ or $z'$ are not defined (because $N = \{x\}$ or
 $N' = \varnothing$), we may suppose them equal to $x$.

 Pick any $y \in F$. Then $\BD(x',y)< \BD(y,x) \leq \BD(x',x)$. Thus we can also use
 \Cref{LEMMA_3Points} and get similarly that there exist points $y, y'\in F$ (where
 $y \in F$ with $\BD(x',y)$ maximal, $F'=\J(x',y)$, and $y'\in F''=F\setminus F'$
 with $\BD(x',y')$ maximal) such that
 \begin{itemize}[label=-]
 \item[$\bullet$] either $\J(x',y') = \arc{y'x'}$ and $\J(x',y) = \arc{x'y}$,
 \item[$\bullet$] or $\J(x',y) = \arc{yx'}$ and $\J(x',y') = \arc{x'y'}$,
 \end{itemize}
 and $F = \J(x',y) \cup \J(x',y')$. From their definitions, it
 immediately follows that the pairs $\{z,z'\}$, $\{y,y'\}$ and the
 partition $\arc{zx} \cup \arc{xz'} \cup \arc{yx'} \cup \arc{x'y'}$
 can be computed in $O(n)$ time.
\end{proof}

Sorting the points of $N', N''$ and $F', F''$ by their distances to
$x$ and to $x'$ takes $O(n\log n)$ time. Then
\cref{LEMMA_fourpartition} allows to partition the compatible circular
order into two ordered sequences. In the first sequence,  $N$ is ordered into
$N = \{x_1,x_2,\ldots,x_k\}$ by taking $N'$ in decreasing order of
distances from $x$, followed by $x$ and then $N''$ in increasing order
of distances from $x$; the second order is the reverse of the first order. Similarly,
$F$ is ordered into $F = \{y_1, y_2, \ldots, y_l\}$ using $F'$, $F''$ and the
distances from $x'$ and its reverse order. This leads to four possibilities to compose the
two ordered (up to reversal) sequences $N = \{x_1,x_2,\ldots,x_k\}$ and
$F = \{y_1, y_2, \ldots, y_l\}$ into a compatible circular order.
Notice that up to symmetry, this reduces to only two possibilities. The
next lemma gives a criterion to decide which one of the two options is
valid.



\begin{lemma}\label{LEMMA_quadruples}
  Let $N = \{ x_1,x_2,\ldots,x_k\}$ and $F = \{ y_1,y_2,\ldots,y_\ell\}$ be the ordered sequences
  defined as above. Let $\beta_1$ and
  $\beta_2$ be the two circular orders on $X$ defined by setting
 \begin{enumerate}[label=(\alph*)]
 \item $x_1 \<_{\beta_1} x_2 \<_1 \ldots \<_{\beta_1} x_k \<_{\beta_1} y_1 \<_{\beta_1} y_2 \<_{\beta_1} \ldots \<_{\beta_1} y_\ell$,
 \item $x_k \<_{\beta_2} x_{k-1} \<_{\beta_2} \ldots \<_{\beta_2} x_1 \<_{\beta_2} y_1 \<_{\beta_2} y_2 \<_{\beta_2} \ldots \<_{\beta_2} y_\ell$.
 \end{enumerate}
 One can decide which of these two circular orders $\beta_1,\beta_2$
 (possibly both) is compatible in $O(n)$ time.
\end{lemma}

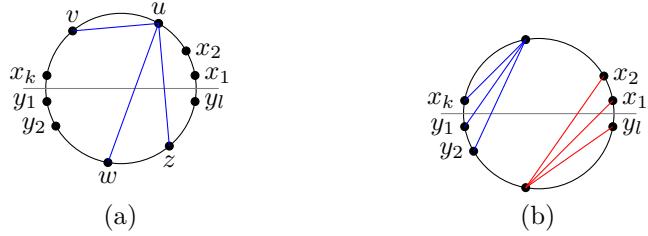
\begin{figure}[htbp]
 \begin{center}
   \begin{tabular}{cp{0.1\textwidth}c}
    \begin{tikzpicture}[x=1cm,y=1cm]
     \draw (0,0) circle[radius=1];
     \draw[gray,thin] (-1.3,0) -- (1.3,0);
     \tkzDefPoint(10:1){x1}
     \tkzDefPoint(30:1){x2}
     \tkzDefPoint(170:1){xk}
     \tkzDefPoint(190:1){y1}
     \tkzDefPoint(210:1){y2}
     \tkzDefPoint(350:1){yl}
     \tkzDefPoint(60:1){u}
     \tkzDefPoint(130:1){v}
     \tkzDefPoint(260:1){w}
     \tkzDefPoint(310:1){z}
     \tkzDrawPoints[vertex](x1,x2,xk,y1,y2,yl,u,v,w,z)
     \node[right=0pt of x1] {$x_1$};
     \node[right=0pt of x2] {$x_2$};
     \node[left=0pt of xk] {$x_k$};
     \node[left=0pt of y1] {$y_1$};
     \node[left=0pt of y2] {$y_2$};
     \node[right=0pt of yl] {$y_l$};
     \node[above=0pt of u] {$u$};
     \node[above=0pt of v] {$v$};
     \node[below=0pt of w] {$w$};
     \node[below=0pt of z] {$z$};
     \foreach \u/\v in {u/v,u/w,u/z} {
         \draw[blue] (\u) -- (\v);
       }
    \end{tikzpicture}
    & &
    \begin{tikzpicture}[x=1cm,y=1cm]
     \draw (0,0) circle[radius=1];
     \draw[gray,thin] (-1.3,0) -- (1.3,0);
     \tkzDefPoint(10:1){x1}
     \tkzDefPoint(30:1){x2}
     \tkzDefPoint(170:1){xk}
     \tkzDefPoint(190:1){y1}
     \tkzDefPoint(210:1){y2}
     \tkzDefPoint(350:1){yl}
     \tkzDefPoint(100:1){u}
     \tkzDefPoint(260:1){v}
     \tkzDrawPoints[vertex](x1,x2,xk,y1,y2,yl,u,v)
     \node[right=0pt of x1] {$x_1$};
     \node[right=0pt of x2] {$x_2$};
     \node[left=0pt of xk] {$x_k$};
     \node[left=0pt of y1] {$y_1$};
     \node[left=0pt of y2] {$y_2$};
     \node[right=0pt of yl] {$y_l$};
     \foreach \u/\v in {u/xk,u/y1,u/y2} {
         \draw[blue] (\u) -- (\v);
       };
     \foreach \u/\v in {v/yl,v/x1,v/x2} {
         \draw[red] (\u) -- (\v);
       };
    \end{tikzpicture}
    \\
    (a) & & (b)
   \end{tabular}
  \end{center}
  \caption{In (a), a configuration occuring in the proof of
    \cref{LEMMA_quadruples}. If $\beta_1$ is not compatible for the
    quadruplet $(u,v,w,z)$, then it is not compatible for the
    quadruplet $(u,x_k,y_1,y_2)$. In (b), an illustration of the two
    families of quadruplets which are enough to check the
    compatibility on.}
 \label{FIG_bad_quadruplets}
\end{figure}

\begin{proof}
  Suppose that $\beta_1$ is not compatible and that is $\beta_2$ is
  compatible. Then there is a quadruplet
  $u \<_{\beta_1} v \<_{\beta_1} w \<_{\beta_1} z$ with
  $\BD(u,w) \leq \min \{\BD(u,v), \BD(u,z)\}$.
  Since $\beta_2$ is a compatible circular order, we must have that
  $(u,v,w,z)$ is one of the four quadruplets
  $(x_i, x_{i'}, y_j, y_{j'})$, $(x_{i'}, y_j, y_{j'}, x_i)$,
  $(y_j, y_{j'},x_i, x_{i'})$, or $(y_{j'},x_i, x_{i'}, y_j)$, with
  $i < i'$ and $j < j'$. Up to symmetry, we may assume the first
  without loss of generality. We may also assume that $j = 1$ and
  $j' = 2$. Indeed, the distances from $x_1$ of the points of $F$,
  from $y_1$ to $y_\ell$, are strictly increasing, then maximal, then
  strictly decreasing, thus by the existence of $j$ and $j'$ the
  increasing sequence is non-empty and $\BD(x_i,y_1) < \BD(x_i,y_2)$.
  Moreover, $\BD(x_i,y_1) \leq \BD(x_i,y_j) < \BD(x_i,x_{i'})$.

  Furthermore, we may also assume that $i' = k$. Indeed, if
  $\BD(x_i,x_k) < \BD(x_i,y_1)$, then
  $\BD(x_i,x_k) < \max \{\BD(x_i,x_i'), \BD(x_i,y_2)\}$, which implies
  that $x_i \<_{\beta_2} x_{i'} \<_{\beta_2} x_{k} \<_{\beta_2} y_2$
  is a quadruplet violating compatibility of $\beta_2$, a
  contradiction. This proves that $\BD(x_i,x_k) \geq \BD(x_i,y_1)$,
  hence $x_i,x_k,y_1,y_2$ is a violating quadruplet. Thus, considering
  the three remaining symmetric cases, we obtain that if there is a
  violating quadruplet, then also there is a violating quadruplet of
  the form $\{x_i,x_k,y_1,y_2\}$ or
  $\{x_i,y_{l-1},y_{l},x_1\}$ for some point $x_i$, or
  $\{y_j,y_\ell,x_1,x_2\}$ or $\{y_j,x_{k-1},x_{k},y_1\}$ for
  some point $y_j$. There are at most $2n$ such quadruplets in
  total, each of them may be checked in $O(1)$ time, summing up to a
  complexity of $O(n)$ time using \cref{ALGO_OA}.
\end{proof}

\subsection{The algorithm} \label{s:algo}

The previous discussion leads to an algorithm for finding a compatible
order, presented in \Cref{ALGO_FCO,ALGO_OA}. The function
$\call{Sort}(x, S)$ sorts $S$ by increasing values of $\BD(x,t)$ for
$t\in S$ (we call this an \emph{$x$-sorting} of the set $S$) and the
function $\call{ReverseSort}(S, x)$ sorts $S$ by decreasing values of
$\BD(x,t)$. The operator $\append$ between two sequences represents
their concatenation into a circular order. Notice that the same
algorithm works for strictly circular and strictly quasi-circular
Robinson dissimilarities, and that the algorithm always outputs an
ordering, which may be arbitrary if the dissimilarity space is not
strictly circular or strictly quasi-circular Robinson.

\begin{algorithm}[htbp]
 \caption{\call{OrdersAgree}}
 \label{ALGO_OA}
 \begin{algorithmic}[1]
\Require{A dissimilarity space $(X,\BD)$, a partition $X = N\cup F$
 with $N = \{x_1,\ldots,x_k\}$ and
 $F = \{y_1,\ldots,y_\ell\}$.}
\Ensure{whether the order $N \append F$ may be compatible based on \Cref{LEMMA_quadruples}.}
\If { $k=1$ or $\ell=1$ }
 \Return true
\EndIf
\ForAll {$i \in \{1,2,\ldots,k\}$}
 \If {not $\sqosR(x_i,x_k,y_1,y_2)$ or not $\sqosR(x_i,y_{\ell-1},y_{\ell-2},x_1)$}
  \Return false
 \EndIf
\EndFor
\ForAll {$i \in \{1,2,\ldots,\ell\}$}
 \If {not $\sqosR(y_i,y_\ell,x_1,x_2)$ or not $\sqosR(y_i,x_{k-1},x_{k-2},y_1)$}
  \Return false
 \EndIf
\EndFor
\Return true
 \end{algorithmic}
\end{algorithm}

\begin{algorithm}[htbp]
 \caption{\call{FindCompatibleOrder}}
 \label{ALGO_FCO}
 \begin{algorithmic}[1]
\Require{A dissimilarity space $(X,\BD)$.}
\Ensure{
 A total ordering
 of $X$, compatible if $(X,\BD)$ is
 (quasi-)circular Robinson.
}
\Let $x \in X$, $x' \in F_x$
\Let $N = \{ u \in X : \BD(u,x) \leq \BD(u,x') \}$
\Let $F = \{ u \in X : \BD(u,x') \leq \BD(u,x) \}$
\If {$N \cap F \neq \varnothing$} \label{step:fcoIfIntersect}
 \Let $y \in N \cup F$ \label{step:fcoDefY}
 \Let $X_1 = \J(x,y) \cup \J(y,x')$ \label{step:fcoDefX1}
 \Let $X_2 = X \setminus X_1 \setminus \{x,x'\} $ \label{step:fcoDefX2}
 \Return $\call{Sort}(x,X_1) \append \call{ReverseSort}(x,X_2)$ \label{step:fcoReturnX1X2}
\Else
 \Let $z = \arg\max_{u \in N} \BD(x,u)$ and $y = \arg\max_{u \in F} \BD(x',u)$ \label{step:fcoDefZ}
 \Let $N' = J(x,z)$ and $F' = J(x',y)$ \label{step:fcoDefN}
 \Let $X_N = \call{ReverseSort}(x,N \setminus N') \append \call{Sort}(x,N')$ \label{step:fcoDefXN}
 \Let $X_F = \call{Sort}(x',F \setminus F') \append \call{ReverseSort}(x',F')$ \label{step:fcoDefXF}
 \If {$\call{OrdersAgree}(X_N,X_F)$} \label{step:fcoIFOA}
  \Return $X_N \append X_F$ \label{step:fcoReturnOA}
 \Else \label{step:fcoElse}
  \Return $X_N \append \call{Reverse}(X_F)$ \label{step:fcoReturnOD}
 \EndIf
\EndIf
 \end{algorithmic}
\end{algorithm}

\begin{theorem}\label{THM_algo}
 \Cref{ALGO_FCO} called to a strictly quasi-circular Robinson or a strictly
 circular Robinson dissimilarity $(X,\BD)$ on $n$ points produces a
 compatible circular order in $O(n \log n)$ time.
\end{theorem}

\begin{proof}
 The correctness of the algorithm follows from
 \Cref{CLAIM_NFarcs,CLAIM_equalcase,LEMMA_fourpartition,LEMMA_quadruples}.
 Namely, \Cref{CLAIM_equalcase} covers the case
 $N\cap F\ne \varnothing$
 (\cref{step:fcoIfIntersect,step:fcoDefY,step:fcoDefX1,step:fcoDefX2,step:fcoReturnX1X2}),
 while \Cref{LEMMA_fourpartition} covers the case
 $N\cap F=\varnothing$
 (\cref{step:fcoDefZ,step:fcoDefN,step:fcoDefXN,step:fcoDefXF,step:fcoIFOA,step:fcoReturnOA,step:fcoElse,step:fcoReturnOD}).
 From these lemmas and \Cref{CLAIM_NFarcs} it follows that the
 circular orders returned in
 \cref{step:fcoReturnX1X2,step:fcoReturnOA,step:fcoReturnOD} are the
 only possible compatible circular orders for $(X,\BD)$. Since
 $(X,\BD)$ is strictly circular Robinson or strictly quasi-circular
 Robinson, we can apply \Cref{LEMMA_quadruples} to deduce that one of
 these circular orders is indeed compatible. The complexity of the
 algorithm is dominated by the time to sort the lists, as every other
 operation can easily be implemented in either constant or linear
 time.
\end{proof}

From \Cref{testing} and \Cref{THM_algo} we immediately obtain the following result:

\begin{corollary}
 For a dissimilarity space $(X,\BD)$ on $n$ points, one can decide in
 optimal $O(n^2)$ time if $(X,\BD)$ is strictly circular Robinson or
 strictly quasi-circular Robinson.
\end{corollary}

The complexity in \Cref{THM_algo} is dominated by the time to sort
the points by their distances to $x$ or $x'$, and is actually
tightly related to the complexity of sorting:

\begin{proposition}
 The problem of sorting a set $Y$ of $n$ distinct integers reduces
 linearly to the problem of finding a compatible circular order for a strictly
 quasi-circular Robinson dissimilarity.
\end{proposition}

\begin{proof}
 Given a set $Y \subseteq \mathbb{N}$, let $X = Y \cup \{z\}$ and let $d$ be
 a dissimilarity on $X$ defined by
 \begin{align*}
  \BD(y,z) &= \Delta + 1 ~~\textrm{for all } y \in Y,\\
  \BD(y,y') &= |y - y'| ~~\textrm{for all } y, y' \in Y,\\
  \BD(z,z) &= 0,
 \end{align*}
 where $\Delta = \max Y - \min Y$. Then it can be
 readily checked that $(X,\BD)$ is a strictly quasi-circular Robinson
 dissimilarity, whose only two compatible orders induce an increasing
 or decreasing ordering of $Y$. This reduction is linear, as long as
 we encode the distance function $d$ as an oracle, to avoid the
 computation of $\Theta(n^2)$ values.
\end{proof}

\subsection{On the number of compatible circular orders}

From \Cref{ALGO_FCO}, we can derive the following result about the
number of compatible orders:

\begin{proposition}\label{THEORM_orders}
 A strict quasi-circular Robinson
 space $(X,\BD)$ has one or two compatible orders and their opposites.
 A strict circular Robinson space has one compatible order and its
 opposite.
\end{proposition}

\begin{proof}
 The first assertion is a direct consequence of \Cref{ALGO_FCO} and
 the proof of \Cref{THM_algo}. Now, let $(X, \BD)$ be a strict circular
 Robinson space with two compatible circular orders $\beta$ and
 $\beta'$. Then $N\cap F=\varnothing$ and the arcs $N$ and $F$ are
 partitioned into $N',N''$ and $F',F''$, respectively (see the proof
 of \Cref{LEMMA_fourpartition}). Then the second compatible order
 $\beta'$ is built from $\beta$ by reversing $N'$ and
 $N''$. If the set $N$ is empty, then this reversal does not change the
 order, thus $\beta'=\beta$. If $F$ is empty, then this reversal
 builds the opposite order of the original one, thus
 $\beta'=\beta^{op}$. So, we can suppose with no loss of generality
 that there exist $y \in N'$ and $z \in F'$ and that the points $y$
 and $z$ are on the same arc $\arc{xx'}$ of $\beta$. 
 The arcs $\arc{xz}$ and $\arc{yx'}$ are
 strictly Robinson, so $\BD(y,z)<\min \{ \BD(x, z), \BD(y, x')\}$. 
 By $\sosR(z, x, y, x')$ applied to $\beta'$, we must have
 $\BD(y,z)>\BD(x,z)$, which is in contradiction with
 $\BD(y,z)<\min \{ \BD(x, z), \BD(y, x')\}$, whence $\beta$ and $\beta'$ cannot be both
 compatible.
\end{proof}


If a strict quasi-circular Robinson space has two compatible orders
and their opposites, then \Cref{ALGO_FCO} yields a bipartition of $X$
into $N \cup F$. Next we prove that this happens exactly when there is
a threshold value that clusters the dissimilarity space into two
cliques:


\begin{proposition}\label{THEOREM_bi_partition}
 Let $(X, \BD)$ be a strict quasi-circular Robinson space. Then $(X,\BD)$
 admits two compatible orders and their opposites if and only if
 there exists a partition $X = N \cup F$ with $|N|, |F| > 1$ and
 $\delta \in \mathbb{R}^+$ such that for all $u, v \in X$, we have
 $\BD(u,v) > \delta$ if and only if $|\{u,v\} \cap N| = 1$.
\end{proposition}

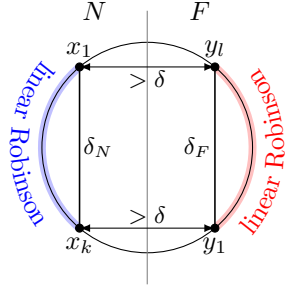
\begin{figure}[htbp]
 \begin{center}
  \begin{tikzpicture}[x=1.4cm,y=1.4cm,>=latex]
   \draw[nearly transparent,line width=3pt,blue] (130:1)
   arc[radius=1,start angle=130,end angle=230] (230:1);
   \draw[nearly transparent,line width=3pt,red] (-50:1)
   arc[radius=1,start angle=-50,end angle=50] (50:1);
   \draw (0,0) circle[radius=1];
   \tkzDefPoint(130:1){x1}
   \tkzDefPoint(50:1){yl}
   \tkzDefPoint(230:1){xk}
   \tkzDefPoint(310:1){y1}
   \tkzDrawPoints[vertex](x1,xk,y1,yl)
   \tkzDrawSegments(x1,xk y1,yl)
   \tkzDefMidPoint(x1,xk)\tkzGetPoint{dn}
   \tkzDefMidPoint(y1,yl)\tkzGetPoint{df}
   \tkzDefMidPoint(x1,yl)\tkzGetPoint{mt}
   \tkzDefMidPoint(xk,y1)\tkzGetPoint{mb}
   \node[right=-2pt of dn] {\small$\delta_N$};
   \node[left=-2pt of df] {\small$\delta_F$};
   \node[above=0pt of x1] {$x_1$};
   \node[below=0pt of xk] {$x_k$};
   \node[below=0pt of y1] {$y_1$};
   \node[above=0pt of yl] {$y_l$};
   \draw (-0.5,1.3) node {$N$};
   \draw (0.5,1.3) node {$F$};
   \draw[gray,thin] (0,-1.3) -- (0,1.3);

   \path[postaction={decorate,decoration={text along path,text color=blue,text={linear Robinson}}}]
   (140:1.3) arc[radius=1.3,start angle=140,end angle=220] (220:1.3);
   \path[postaction={decorate,decoration={text along path,text color=red,text={linear Robinson}}}]
   (-40:1.3) arc[radius=1.3,start angle=-40,end angle=40] (40:1.3);

   \draw[<->] (x1) -- (yl);
   \node[below=-2pt of mt] {\small$> \delta$};
   \draw[<->] (xk) -- (y1);
   \node[above=-2pt of mb] {\small$> \delta$};
  \end{tikzpicture}
 \end{center}
 \caption{The structure of a strictly quasi-circular Robinson
  space with two non-opposite compatible orders, with
  $\delta = \max \{\delta_N, \delta_F\}$, as shown by
  \Cref{THEOREM_bi_partition}. $N$ and $F$ have diameters $\delta_N$
  and $\delta_F$ respectively, and all pairs between $N$ and $F$
  have distance greater than $\delta$. The proof that $N$
  (symmetrically, $F$) are linear Robinson follows easily from
  $\sqosR(x_{i_1},x_{i_2},x_{i_3},y_1)$ and
  $\sqosR(x_{i_3},y_1,x_{i_1},x_{i_2})$.}
 \label{FIG_two_orders}
\end{figure}

\begin{proof}
 Suppose first that $(X,\BD)$ admits two compatible orders and their
 opposites. By \Cref{LEMMA_fourpartition,LEMMA_quadruples}, there is
 a bipartition $N \cup F$ with $N = \{x_1,x_2,\ldots,x_k\}$ and
 $F = \{y_1,y_2,\ldots,y_\ell\}$, such that the compatible orders are
 $\beta$ (given by $N \append F$), $\beta'$ (given by
 $N \append \call{Reverse(F)}$), and their reverses. Notice that
 $k > 1$ and $\ell > 1$. Let $\delta_N = \BD(x_1,x_k)$ and
 $\delta_F = \BD(y_1,y_\ell)$. Then for any distinct
 $j, j' \in \{1,2,\ldots,\ell\}$, $\sqosR(x_k,y_j,y_{j'},x_1)$ (in
 $\beta$) and $\sqosR(x_k,y_{j'},y_{j},x_1)$ (in $\beta'$) we have:
 \begin{align*}
  \BD(x_k,y_{j'}) &> \min \{ \BD(x_k,x_1), \BD(x_k,y_j)\}, \\
  \BD(x_k,y_j) &> \min \{ \BD(x_k,x_1), \BD(x_k,y_{j'})\}.
 \end{align*}
 Thus $\delta_N = \BD(x_1,x_k) < \min \{ \BD(x_k,y_j), \BD(x_k,y_{j'})\}$.
 Analogously, $\delta_N < \min \{ \BD(x_1.y_j), \BD(x_1,y_{j'})\}$. Then for
 any $i \in \{2,3,\ldots,k-1\}$, by $\sqosR(y,x_1,x_i,x_k)$,
 $\BD(x_i,y) > \min \{\BD(y,x_1),\BD(y,x_k)\} > \delta_N$. This proves that
 $\min \{\BD(x,y) : x \in N, y \in F\} > \delta_N$.

 Consequently, for any $y \in F$ and $i \in \{1,2,\ldots, k-1\}$, by
 $\sqosR(x_k,y,x_1,x_i)$,
 $\delta_N = \BD(x_k,x_1) > \min \{\BD(x_k,y),\BD(x_k,x_i)\}$, which
 implies that $\BD(x_i,x_k) < \delta_N$. For
 $j \in \{i+1,i+2,\ldots,k-1\}$, by $\sqosR(x_i,x_j,x_k,y))$,
 $\BD(x_i,x_k) > \min \{\BD(x_i,x_j), \BD(x_i,y)\}$, which implies that
 $\BD(x_i,x_j) < \delta_N$, hence
 $\max \{\BD(u,v) : u, v \in N\} = \delta_N$. Analogously,
 we have $\max \{\BD(u,v) : u,v \in F\} = \delta_F$ and
 $\min \{\BD(x,y) : x \in N, y \in F\} > \delta_F$. Thus taking
 $\delta = \max \{\delta_N,\delta_F\}$ proves the assertion.

 Conversely, suppose that $(X,\BD)$ is a strictly quasi-circular
 Robinson space admitting such a bipartition $X = N \cup F$. Clearly
 $N$ and $F$ are balls of radius $\delta$ and thus, in any compatible
 order, by \Cref{PROPOSITION_ball}, $N$ and $F$ are arcs. Let
 $x_1 \< x_2 \< \ldots \< x_k \< y_1 \< y_2 \< \ldots \< y_\ell$ be a
 compatible order $\beta$, with $N = \{ x_1, x_2,\ldots, x_k\}$ and
 $F = \{ y_1, y_2, \ldots, y_\ell\}$. Then, we can check that for any
 quadruplet $u \< v \< w \< t$ of the circular order $\beta'$ induced by
 $N \append \call{Reverse}(F)$, $\sqosR(u,v,w,t)$ holds. Indeed, the
 only nontrivial case (where the circular order is distinct for
 $\beta$ and $\beta'$ up to reversal) is when $u, v \in N$ and
 $w, t \in F$ (up to symmetry). In that case, we have
 $\BD(u,w) > \beta \geq \BD(u,v) \geq \min \{\BD(u,s),\BD(u,v)\}$, that is
 $\sqosR(u,v,w,t)$. This implies that $\beta'$ is also compatible. Since $k, \ell > 1$,
 $\beta$ and $\beta'$ are not the reverse of each other, proving the
 proposition.
\end{proof}

\section{Conclusion} \label{s:concl}

In this paper, we presented a very simple algorithm which solves the
strict quasi-circular and strict
circular seriation problems in optimal $O(n^2)$
time.
%
%
Notice that the $O(n^2)$ time is entirely due to the verification of
the result, while the computation of a compatible circular order (the
main part of the algorithm) is in $O(n\log n)$ time. In addition,
using the algorithm we proved some structural properties of strictly
quasi-circular and strictly circular Robinson spaces. We also proved
that any pre-circular Robinson space is circular Robinson, a result
which can find further applications. As we already noticed in the
introduction, designing an algorithm which solves the circular
seriation problem in $O(n^2)$ (or even in $O(n^2\log n)$ time) is an
interesting \emph{open question}. Designing approximation algorithms
for fitting a dissimilarity by a circular Robinson dissimilarity is
another \emph{open problem}.

\subsection*{Acknowledgement}
We would like to acknowledge the referees for their careful reading of the manuscript and useful suggestions and comments. This research was supported in part by ANR project DISTANCIA
(ANR-17-CE40-0015) and has received funding from Excellence Initiative
of Aix-Marseille - A*MIDEX (Archimedes Institute AMX-19-IET-009), a
French ”Investissements d’Avenir” Programme.

\bibliographystyle{siamplain}
\bibliography{citations.bib}

\end{document}